\documentclass[12pt, a4paper]{article}

\usepackage[utf8]{inputenc}
\usepackage[T1]{fontenc}
\usepackage{lmodern}
\usepackage{microtype}
\usepackage{amsmath, amssymb, amsfonts}
\usepackage{bm}
\usepackage{graphicx}
\usepackage{xcolor}
\usepackage{booktabs}
\usepackage{multirow}
\usepackage{siunitx}
\usepackage{natbib}
\usepackage{hyperref}
\usepackage{geometry}
\usepackage{setspace}
\usepackage{caption}
\usepackage{subcaption}
\usepackage{enumitem}
\usepackage[labelfont=bf]{caption}
\usepackage{comment}

\hypersetup{
    colorlinks = true,
    linkcolor  = blue!60!black,
    citecolor  = blue!60!black,
    urlcolor   = blue!70!black
}

\newcommand{\Msun}{M_{\odot}}
\newcommand{\Mjup}{M_{\rm Jupiter}}
\newcommand{\ajup}{a_{\rm Jupiter}}

\newcommand{\au}{\,\text{au}}
\newcommand{\vkick}{v_{\rm kick}}

\newcommand{\vJ}{v_J}
\newcommand{\RH}{R_{\rm H}}
\newcommand{\vH}{v_{\rm H}}
\newcommand{\TJ}{T_J}

\begin{document}

\title{%
    \textbf{\large Kick Velocities and Mass Function of Free-Floating Planets
    from Dynamical Ejection in Hierarchical Three-Body Systems}
}

\author{
    Hugh Kramer\,$^{1}$\thanks{Corresponding author: hugh\_kramer27@milton.edu} and 
    Stefano Profumo\,$^{2,3}$\thanks{profumo@ucsc.edu}
    \\[1.2ex]
    \small $^{1}$Milton Academy, Milton, MA, USA \\
    \small $^{2}$Department of Physics, University of California Santa Cruz\\
    \small $^{3}$Santa Cruz Institute for Particle Physics, 1156 High St., Santa Cruz, CA 95064, USA
}

\date{}

\maketitle

\begin{abstract}
\noindent Free-floating planets (FFPs), also known as rogue planets, are sub-stellar objects that travel through the Galaxy unbound to any host star. Their velocity distribution carries information about the dynamical channel that released them from their birth systems, while their mass distribution encodes the underlying abundance of planets available for ejection. We present a suite of direct $N$-body simulations of hierarchical three-body systems consisting of a Solar-mass host star, a massive giant perturber, and a lighter planet treated in the restricted three-body regime. We vary the mass of the ejected planet, the mass of the giant perturber, the light planet's semi-major axis, and the eccentricities of both planets, measuring the asymptotic ejection velocity $v_\infty$ and the finite-radius ejection speed $v_{\rm kick}$ at the first accepted ejection output. The ejected body's mass has little effect on the outcome over four orders of magnitude, confirming the test-particle limit. By contrast, the giant-planet mass sets the ejection scale and time, following the secular scaling $t_{\rm ejec}\propto M_J^{-1}$. The eccentricities mainly affect the high-velocity tail rather than the median: a higher light-planet eccentricity extends the kick ceiling to $\sim23$ km/s, while a highly eccentric giant can yield rare kicks near $80$ km/s via pericentre-enhanced slingshot encounters. We interpret these trends with a semi-analytic framework based on Hill-scale scattering, the Tisserand parameter, and an eccentricity-dependent upper envelope for slingshot energy exchange, and discuss how the ejection velocities map onto the Galactic FFP velocity dispersion and how the mass function sets the microlensing timescale distribution relevant for Roman and Euclid.
\end{abstract}

\newpage
\tableofcontents
\newpage

\section{Introduction}
\label{sec:intro}

Free-floating planets (FFPs)—planetary-mass objects unbound to any stellar host—travel
through the Galactic field on purely gravitational trajectories set at the moment of
their ejection or formation. Because they carry a kinematic fossil record of the
dynamical environment in which they originated, the velocity and mass distributions of
the FFP population provide a direct probe of both planetary-system demographics and the
efficiency of planet-forming processes across cosmic time.

The primary observational channel for detecting FFPs is gravitational microlensing.
Because FFPs are intrinsically dark, they reveal themselves only through the brief,
symmetric brightening they produce as they transit the line of sight to a background
source star. The Einstein-ring crossing timescale $t_{\rm E}$ scales as the square root
of the lens mass, placing planetary-mass objects in the regime $t_{\rm E} \lesssim
1$\,day. Early evidence for a substantial Jupiter-mass FFP population came from
\citep{Sumi2011}, who found an excess of short-timescale events in the Microlensing
Observations in Astrophysics (MOA) survey inconsistent with known stellar populations.
Subsequent analyses of Optical Gravitational Lensing Experiment (OGLE) data confirmed
sub-day events consistent with Earth- to Jupiter-mass lenses
\citep{Mroz2017,Mroz2019,Mroz2020}; most recently, \citep{Mroz2020} identified
ultrashort events suggestive of a sub-Earth-mass component, raising the possibility that
the mass spectrum of ejected objects extends well into the rocky-planet regime.
Complementary detections have come from near-infrared imaging of young stellar
associations, where residual formation heat renders isolated planetary-mass objects
accessible for ages up to a few tens of Myr \citep{Luhman2012,Scholz2012,Miret-Roig2022}.
The Upper Scorpius association alone has yielded dozens of planetary-mass candidates
down to a few Jupiter masses \citep{Lodieu2013,Pearson2023}. These directly imaged
objects probe the high-mass end of the FFP spectrum and may include a contribution
from in-situ formation via core collapse \citep{Boss2001,Padoan2004}; the lower-mass
microlensing population, however, is almost certainly dominated by objects dynamically
ejected from planetary systems.

Several production channels have been proposed. Dynamical ejection from multi-planet
systems is arguably the most efficient: close encounters drive one body onto a hyperbolic
orbit while the other settles into a tighter, more eccentric configuration
\citep{Rasio1996,Weidenschilling1996,Lin1997}. $N$-body studies show that
planet--planet scattering is nearly ubiquitous in systems that go unstable on Gyr
timescales \citep{Chatterjee2008,Juric2008,Raymond2010}, and that a significant fraction
produce at least one unbound planet per scattering episode \citep{Veras2009,Barclay2017}.
Stellar flybys in dense cluster environments can strip outer planets from their hosts
during the first few hundred Myr of a star's life \citep{Malmberg2007,Parker2012,Cai2019},
and disc photoevaporation or tidal disruption may unbind planets still forming at the
time of a close encounter \citep{Adams2003}. Finally, in-situ gravitational fragmentation
of dense molecular filaments can produce isolated planetary-mass objects with no prior
bound phase \citep{Padoan2004,Whitworth2007}. Of these channels, dynamical ejection is
uniquely amenable to systematic numerical study, because the outcome maps onto a
well-defined and controllable set of initial conditions, and because the resulting kick
velocity is in principle directly observable through the proper-motion distribution of
imaged FFPs or the microlensing parallax signal \citep{Gould2022,Ban2020}.

The simplest system in which dynamical ejection can occur is the hierarchical three-body
problem: a host star with two planets, where a massive outer perturber drives the inner,
lighter planet to instability through a combination of secular forcing, resonance crossing,
and eventual close encounters. This configuration captures the essential physics while
remaining tractable to large parameter surveys. Previous studies have mapped
instability timescales across subsets of the relevant parameter space
\citep{Gladman1993,Chambers1996,Marzari2002,Pfyffer2015}, but a comprehensive
characterisation of the resulting \emph{kick velocity} distribution has not been
undertaken.

The asymptotic ejection velocity is the observable of central importance. If
$v_\infty$ is small relative to the local Galactic velocity dispersion
($\sigma \approx 30$--$50\,\text{km\,s}^{-1}$ for thin-disc stars), the FFP
population is kinematically indistinguishable from its parent stellar population. If it
is large, FFPs form a dynamically hotter component with a distinct proper-motion
signature. A speed measured at finite radius also contains the contribution required to
climb out of the host-star potential and must not be identified with $v_\infty$
(Section~\ref{ssec:vesc}). Mapping the full asymptotic-velocity distribution—including
its high-velocity tail—is therefore essential for predicting what \textit{Gaia},
\textit{Roman}, and \textit{Euclid} will observe \citep{Penny2019,Johnson2020}.

In this paper we present a large suite of direct $N$-body simulations of hierarchical
three-body systems to address three questions: (i) what is the distribution of kick velocities
imparted to ejected planets, and how does it depend on system architecture; (ii) what
is the distribution of ejection timescales, and which parameters drive it; and (iii) how
do these results constrain the mass function and velocity dispersion of the present-day
Galactic FFP population. The present simulation grid fixes the host star to one Solar
mass and the giant planet's semi-major axis to a Jupiter analog, and then varies the
light-planet mass, giant-planet mass, light-planet semi-major axis, and the two orbital
eccentricities. This controlled grid isolates the physical role of each parameter before
the results are folded into a broader population model.

The paper is organised as follows. Section~\ref{sec:theory} develops the analytical
framework governing ejection energetics, secular evolution, and escape velocities.
Section~\ref{sec:methods} describes the numerical setup, parameter grid, and
integration strategy. Section~\ref{sec:results} presents the simulation results,
organised by parameter: ejected-planet mass, giant-planet mass, semi-major axis,
and orbital eccentricities of both bodies.
Section~\ref{sec:massfunc} derives implications for the observable FFP population.
Section~\ref{sec:discussion} compares with prior work, discusses caveats, and outlines
observational predictions. Section~\ref{sec:conclusions} summarises our conclusions.

\section{Analytical Framework}
\label{sec:theory}

Before describing the numerical setup we review the analytical structure that
underlies ejection in hierarchical three-body systems. This serves two purposes:
it provides physical intuition for the scaling relations found in the simulations,
and it supplies the analytic expressions against which the numerical distributions
can be compared.

\subsection{Energy budget of a three-body ejection}
\label{ssec:energy}

Consider a hierarchical three-body system consisting of a host star of mass $M_\star$,
an outer giant planet of mass $M_J$ on an orbit of semi-major axis $a_J$, and an inner
planet of mass $m_p \ll M_J$ on an orbit of semi-major axis $a_p < a_J$. Because
$m_p \ll M_J$, the inner planet can be treated as a test particle: its gravity does
not significantly perturb either the star or the giant. This is the
\emph{restricted three-body problem}, and our simulations confirm that the ejection
statistics are insensitive to $m_p$ across four orders of magnitude in mass
(Section~\ref{ssec:res_mlight}).

In the unperturbed limit, the specific orbital energy of the inner planet in the field
of the star is
\begin{equation}
    \epsilon_p = -\frac{G M_\star}{2 a_p}\,.
\label{eq:ep}
\end{equation}
This is negative for a bound orbit. At ejection the planet escapes to infinity with
asymptotic speed $v_\infty$, so the specific energy becomes
$\epsilon_\infty = \tfrac{1}{2}v_\infty^2 > 0$. The encounter with the giant must
therefore supply an energy increment
\begin{equation}
    \Delta\epsilon = \frac{1}{2}v_\infty^2 + \frac{G M_\star}{2 a_p}\,.
\label{eq:Depsilon}
\end{equation}
A planet with a larger semi-major axis (weaker binding) requires a smaller
$\Delta\epsilon$ to be ejected.

\subsection{The gravitational slingshot and the maximum kick velocity}
\label{ssec:slingshot}

The mechanism that delivers $\Delta\epsilon$ during a close encounter is the
gravitational slingshot. It is most transparent in the rest frame of the giant planet.

\paragraph{Encounter velocity.}
The inner planet approaches the giant with relative velocity
$\bm{u} = \bm{v}_p - \bm{v}_J$,
where $\bm{v}_p$ is the heliocentric velocity of the inner planet and $\bm{v}_J$ is
that of the giant. In the giant's frame the encounter is approximately elastic
($|\bm{u}_{\rm out}| = |\bm{u}_{\rm in}| = u$), so the giant mainly deflects the
direction of $\bm{u}$ without changing its magnitude. Transforming back to the
heliocentric frame, the change in specific energy is
\begin{equation}
    \Delta\epsilon = \bm{v}_J \cdot (\bm{u}_{\rm out} - \bm{u}_{\rm in})\,.
\label{eq:slingshot}
\end{equation}
The maximum possible energy kick, obtained when $\bm{u}$ is reversed, is
\begin{equation}
    \Delta\epsilon_{\rm max} \approx 2\,v_J\,u\,,
\label{eq:Depsilon_max}
\end{equation}
which sets the upper envelope of the ejection-speed distribution.

\paragraph{Encounter speed as a function of eccentricity.}
For a coplanar orbit with semi-major axis $a_p$ and eccentricity $e_p$, the vis-viva
equation gives the total heliocentric speed at $r = a_J$:
\begin{equation}
    v_p^2 = G M_\star\left(\frac{2}{a_J} - \frac{1}{a_p}\right).
\label{eq:visviva}
\end{equation}
The tangential component at that radius follows from angular-momentum conservation:
\begin{equation}
    v_t = \frac{\sqrt{G M_\star a_p(1-e_p^2)}}{a_J}
        = \vJ \sqrt{x(1-e_p^2)}\,,
\label{eq:vt}
\end{equation}
where $x \equiv a_p/a_J$ and $\vJ = \sqrt{G M_\star / a_J}$ is the giant's circular
speed. The encounter speed then follows from
$u^2 = v_p^2 - v_t^2 + (v_t - \vJ)^2$, giving
\begin{equation}
    \frac{u^2}{\vJ^2} = 3 - \frac{1}{x} - 2\sqrt{x(1-e_p^2)}\,.
\label{eq:uenc}
\end{equation}
Equation~\eqref{eq:uenc} is exact for coplanar crossings of the giant's orbit. It shows
that $u$ increases monotonically with $e_p$ at fixed $x$: higher eccentricity
reduces the tangential speed $v_t$ and raises the radial speed, increasing the
velocity mismatch with the giant.

\paragraph{Maximum ejection speed.}
Combining Equations~\eqref{eq:Depsilon}, \eqref{eq:Depsilon_max}, and \eqref{eq:uenc},
the approximate upper envelope of the ejection-speed distribution is
\begin{equation}
    \frac{v_{\infty,\rm max}^2}{\vJ^2}
    \approx
    4\,\frac{u}{\vJ} - \frac{1}{x}
    =
    4\left[3 - \frac{1}{x} - 2\sqrt{x(1-e_p^2)}\right]^{1/2}
    - \frac{1}{x}\,.
\label{eq:vinf_max}
\end{equation}
This formula does not predict individual ejection speeds; it predicts the envelope that
the distribution cannot exceed. For the fiducial configuration $x = 0.8$
(i.e.\ $a_p = 0.8\,a_J$) and a Jupiter-analog ($\vJ \approx 13.1\,\text{km\,s}^{-1}$)
around a Solar-type host, Equation~\eqref{eq:vinf_max} gives:

\begin{center}
\begin{tabular}{ccc}
\toprule
$e_p$ & $u/\vJ$ & $v_{\infty,\rm max}\;[\text{km\,s}^{-1}]$ \\
\midrule
0.4 & 0.33 &  3.7 \\
0.6 & 0.56 & 13.2 \\
0.8 & 0.82 & 18.7 \\
0.9 & 0.99 & 21.5 \\
\bottomrule
\end{tabular}
\end{center}

\noindent
This reproduces the key feature of the simulated kick-velocity distributions: the
upper cutoff rises steeply with $e_p$, while the bulk of the distribution
(controlled by typical rather than optimal encounters) shifts only modestly. The
numerical results confirm this; maximum kick velocities grow from
${\sim}7.5\,\text{km\,s}^{-1}$ at $e_p = 0$ to ${\sim}23\,\text{km\,s}^{-1}$ at
$e_p = 0.9$ in the inner configuration (Section~\ref{ssec:res_ep}), in reasonable
agreement with the analytic envelope.

\paragraph{Deflection angle.}
The amount by which the giant deflects $\bm{u}$ depends on the impact parameter $b$:
\begin{equation}
    \theta = 2\arctan\!\left(\frac{G M_J}{b\,u^2}\right).
\label{eq:deflection}
\end{equation}
At fixed $b$, larger $u$ gives a smaller deflection angle. There is therefore a
competition: high eccentricity raises $u$ and thus the available slingshot energy, but
it simultaneously reduces the deflection efficiency. The largest kicks arise from a
compromise between sufficient energy and sufficient deflection.

\subsection{Hill-scale scattering and the median ejection speed}
\label{ssec:hill}

While the maximum kick velocity is set by the eccentricity-dependent slingshot
envelope, the \emph{median} ejection speed is controlled by the typical energy
exchanged in encounters that occur within the giant's gravitational sphere of influence.
The characteristic scale of that region is the Hill radius,
\begin{equation}
    \RH = a_J\left(\frac{M_J}{3 M_\star}\right)^{1/3}.
\label{eq:RH}
\end{equation}
For a Jupiter-mass perturber ($M_J/M_\star \simeq 10^{-3}$) at $a_J = 5.2\au$,
$\RH \simeq 0.07\,a_J \simeq 0.36\au$.
The corresponding Hill velocity is
\begin{equation}
    \vH = \Omega_J \RH = \vJ\left(\frac{M_J}{3 M_\star}\right)^{1/3},
\label{eq:vH}
\end{equation}
where $\Omega_J = \vJ/a_J$ is the giant's mean motion. For the same Jupiter analog,
$\vH \approx 0.9\,\text{km\,s}^{-1}$.
Typical ejections from the Hill sphere, accumulated over multiple encounters, produce
asymptotic speeds of order a few $\vH$:
\begin{equation}
    \langle v_\infty \rangle \sim \text{few} \times \vH
    \sim 3\text{--}6\,\text{km\,s}^{-1}\,.
\label{eq:vmed}
\end{equation}
This estimate is largely insensitive to the initial eccentricity of the inner planet,
because typical encounters probe the Hill sphere regardless of $e_p$. It is, however,
sensitive to $M_J$ through Equation~\eqref{eq:vH}: $\vH \propto M_J^{1/3}$, so the
median kick speed is expected to scale as $M_J^{1/3}$.
A concise summary of the two regimes is:
\begin{align}
    \langle v_\infty \rangle &\sim \text{few} \times \vH \propto M_J^{1/3}
    \quad \text{(median; Hill-scale scattering)}\,, \label{eq:median_scaling}\\
    v_{\infty,\rm max} &\propto u(e_p)^{1/2} \propto [1-(1-e_p^2)^{1/2}]^{1/4}
    \quad \text{(tail; eccentricity-dependent slingshot)}\,. \label{eq:tail_scaling}
\end{align}

\subsection{The Tisserand parameter as a conserved diagnostic}
\label{ssec:tisserand}

A useful approximate invariant in the restricted three-body problem is the Tisserand
parameter with respect to the giant planet,
\begin{equation}
    \TJ = \frac{a_J}{a_p} + 2\cos i\sqrt{\frac{a_p}{a_J}(1-e_p^2)}\,,
\label{eq:TJ}
\end{equation}
which is approximately conserved during encounters as long as the giant is the dominant
perturber \citep{Murray1999}. For coplanar prograde orbits ($i = 0$, $\cos i = 1$)
the Tisserand parameter reduces to
\begin{equation}
    \TJ = \frac{1}{x} + 2\sqrt{x(1-e_p^2)}\,,
\label{eq:TJ_coplanar}
\end{equation}
and the encounter speed from Equation~\eqref{eq:uenc} can be written as
\begin{equation}
    \frac{u^2}{\vJ^2} = 3 - \TJ\,,
\label{eq:u_TJ}
\end{equation}
so that $u$ is uniquely determined by $\TJ$ in the coplanar case. Equation~\eqref{eq:u_TJ}
makes clear that low-$\TJ$ orbits (high eccentricity, or outer orbits at large $a_p/a_J$)
arrive at the encounter with higher relative speeds, and hence have access to larger
slingshot energy kicks. The condition for the encounter even to be possible is
$u^2 \geq 0$, i.e.\ $\TJ \leq 3$, which is the standard criterion for orbit-crossing
in the restricted problem. For the circular-giant, nearly coplanar problem, the Tisserand parameter therefore
provides a compact way of organising the encounter kinematics: lower $\TJ$ corresponds
to a larger relative encounter speed and a higher slingshot-energy ceiling. We use this
relation here as an interpretive diagnostic; a direct correlation analysis based on
$\TJ$ evaluated immediately before the first close encounter is left for future work.

\subsection{Secular evolution and the path to ejection}
\label{ssec:secular}

In the hierarchical limit $a_p \ll a_J$, the orbit-averaged (secular) interaction
between the inner and outer planets drives slow precession of the orbital elements
of both bodies while conserving their semi-major axes to first order
(Laplace--Lagrange theory). The secular timescale for the inner planet is
\begin{equation}
    t_{\rm sec} \sim \frac{M_\star}{M_J}\,\frac{P_J^2}{P_p}\,,
\label{eq:tsec}
\end{equation}
where $P_J$ and $P_p$ are the orbital periods of the giant and the inner planet,
respectively. When the mutual inclination $i$ is large enough
($\cos^2 i < 3/5$, i.e.\ $i \gtrsim 39^\circ$), the Kozai--Lidov (KL) mechanism
\citep{Kozai1962,Lidov1962} drives large-amplitude coupled oscillations in inclination
and eccentricity, with the inner orbit reaching a maximum eccentricity
\begin{equation}
    e_{\rm max} = \sqrt{1 - \frac{5}{3}\cos^2 i}\,.
\label{eq:emax_KL}
\end{equation}
At high $e_{\rm max}$, the inner planet's apocentre
$Q_p = a_p(1+e_p)$ can expand into the giant's strong-encounter zone,
$Q_p \gtrsim a_J$, triggering the slingshot mechanism described above. For an
initially outer light planet, the corresponding crossing condition instead involves
the contraction of its pericentre, $q_p = a_p(1-e_p) \lesssim a_J$.

The present simulations are nearly coplanar, with initial inclinations below
$3^\circ$, and therefore do not activate the Kozai--Lidov mechanism. The discussion
above describes an additional injection channel relevant to more highly inclined
systems rather than the mechanism operating in the present simulation grid.

Superimposed on the secular evolution is the phenomenon of mean-motion resonance (MMR)
overlap. When the ratio of orbital frequencies $\Omega_p / \Omega_J$ is close to
a rational number $p{:}q$, the corresponding MMR can trap the inner orbit and drive
chaotic diffusion in eccentricity once the libration widths of adjacent resonances
overlap (Chirikov criterion). The critical separation below which first-order MMRs
overlap is approximately
\begin{equation}
    \frac{\Delta a}{a_J} \lesssim 1.5\left(\frac{M_J}{M_\star}\right)^{2/7}\,,
\label{eq:chirikov}
\end{equation}
\citep{Wisdom1980,Duncan1989}. Systems with $a_p/a_J$ placing the inner planet
inside this zone are expected to go unstable on timescales much shorter than
$t_{\rm sec}$.

\subsection{Local escape speed and the relation between finite-radius and asymptotic velocities}
\label{ssec:vesc}

The local escape speed is the speed required for zero orbital energy at a specified
radius; it is not a lower bound on the asymptotic speed $v_\infty$. Neglecting the
giant once the planet is sufficiently far from the encounter, conservation of specific
energy gives
\begin{equation}
    \frac{1}{2}v^2(r)-\frac{G M_\star}{r}=\frac{1}{2}v_\infty^2\,,
\end{equation}
or equivalently
\begin{equation}
    v^2(r)=v_\infty^2+v_{\rm esc}^2(r)\,,
    \qquad
    v_{\rm esc}(r)=\sqrt{\frac{2 G M_\star}{r}}
    \approx 42.1\,\text{km\,s}^{-1}
    \left(\frac{M_\star}{\Msun}\right)^{1/2}
    \left(\frac{r}{1\,\text{au}}\right)^{-1/2}.
\label{eq:vesc}
\end{equation}
A marginally unbound planet may therefore have $v_\infty\rightarrow0$ while its local
speed approaches $v_{\rm esc}(r)$. For the fiducial Jupiter analog
($M_\star=1\,\Msun$, $a_J=5.2\au$), $v_{\rm esc}(a_J)\approx18.5\,\text{km\,s}^{-1}$,
but the finite-radius quantity $\vkick$ is recorded much farther from the star, after
the accepted ejection criterion has been reached. Equation~\eqref{eq:vesc} explains why
a finite-radius ejection speed can exceed $v_\infty$: the difference reflects the
remaining stellar potential energy, not an additional dynamical impulse. Galactic
kinematics must therefore be constructed from $v_\infty$, which is the velocity retained
relative to the host after the planet has escaped.

\section{Numerical Methods}
\label{sec:methods}

\subsection{Integrator and code description}
\label{ssec:integrator}

All simulations were performed using the \textsc{rebound} N-body integration package. The IAS15 integrator was chosen for its 15th-order adaptive timestepping. Unlike fixed-step integrators, IAS15 automatically adjusts its step size to maintain high accuracy during close encounters, where gravitational forces vary on timescales far shorter than the orbital period. Given the repeated occurrence of close encounters in three-body systems and the statistical nature of this study, IAS15's ability to keep systematic errors below machine precision was necessary.

Although IAS15 maintained near machine-precision energy conservation throughout most simulations, a small fraction of runs exhibited anomalously large energy errors. These outlier runs, where the fractional energy error $\Delta E/E_0 > 10^{-6}$, were excluded from analysis.

We used \textsc{rebound} version 5.0.1 with an IAS15 tolerance of $10^{-9}$. For each run, we recorded the integration state every $2\%$ of the shortest orbital period in the system at initialization. If no ejection was confirmed before $t_{\max} = 10^{6}~\mathrm{yr}$, we ended the simulation and flagged the run as a non-ejection.

\subsection{Initial conditions and parameter grid}
\label{ssec:grid}

Each discrete parameter configuration was simulated with $N = 10{,}000$ independent realizations. A shared set of random seeds was used across all parameter configurations, so that for a given seed the mean anomalies of all bodies, drawn uniformly from $[0, 2\pi]$, and the inclinations, drawn uniformly from $[0^\circ, 3^\circ]$, are identical across figures. This ensures that differences between parameter configurations reflect only the varied parameter, and not differences in initial orbital phases. All other orbital angles are left at their \textsc{rebound} defaults. The mass of the host star is fixed at $M_\star = 1\,\mathrm{M_\odot}$. The semi-major axis of the giant planet is initialized by $a_J=5.2$ AU ($\ajup$). Unless otherwise varied, eccentricities are set to $0$.

\begin{table}[htbp]
\centering
\caption{Parameter grid explored across all simulation sets.}
\resizebox{\textwidth}{!}{%
\begin{tabular}{lllll}
\toprule
Parameter & Symbol & Range & Fixed \\
\midrule
Small planet mass & $m_{p}$ & $10^{-5}$--$10^{-1}\,\Mjup$ & $M_J=10\,\Mjup,\, a_p=0.9\,a_J$\\
Giant planet mass & $M_{J}$ & $1$--$75\,\Mjup$ & $m_p=10^{-3}\,\Mjup,\, a_p=0.9\,a_J$\\
Small planet semi-major axis & $a_p$ & $0.7$--$1.5\,a_J$ & $M_J=10\,\Mjup,\, m_p=10^{-3}\,\Mjup$\\
Small planet eccentricity & $e_p$ & $0.0$--$0.9$ & $a_p \in \{0.8, 1.2\}\,a_J,\, M_J=10\,\Mjup,\, m_p=10^{-3}\,\Mjup$\\
Giant planet eccentricity & $e_J$ & $0.0$--$0.9$ & $a_p \in \{0.8, 1.2\}\,a_J,\, M_J=10\,\Mjup,\, m_p=10^{-3}\,\Mjup$\\
\bottomrule
\end{tabular}%
}
\label{tab:params}
\end{table}

\subsection{Ejection detection and velocity measurement}
\label{ssec:detection}

A particle is flagged as a candidate for ejection if it simultaneously meets the following three criteria:
\begin{itemize}
    \item Its distance from the host star exceeds ten times the maximum semi-major axis in the system.
    \item Its total specific energy is positive.
    \item Its radial velocity relative to the host star is directed outwards.
\end{itemize}
These conditions must persist continuously for ten orbital periods of the longest-period remaining bound body, during which the candidate particle's distance from the star must be increasing on average. If the candidate re-binds at any point during this confirmation window, it is rejected, and the search continues. We define $t_{\rm ejec}$ as the time of the last stored output at which all three candidate criteria are met, immediately before the event passes this confirmation test. The quantity denoted $\vkick$ is the magnitude of the planet's velocity relative to the host star at that same stored output. It is therefore a finite-radius, star-centric ejection speed, not the instantaneous velocity impulse delivered during the close encounter and not the asymptotic speed of the escaped planet.

Once the ejection is confirmed, the simulation continues integrating in steps of the longest remaining bound orbital period until the fractional rate of change of the potential energy of the ejected particle $|\dot{E}_{\rm pot}/E| < 10^{-9}$. At this point, $E$ is taken as the asymptotic value of the particle's energy, free from subsequent interactions with the bound bodies. The particle's ejection velocity at infinity is then computed as $v_{\infty} = \sqrt{2E}$. If convergence is not reached by $3t_{\max}$, the particle's energy at that time is recorded and converted to an approximate asymptotic velocity by the same formula.

\section{Numerical Results}
\label{sec:results}

We present the simulation output organised around three observables that appear in
each figure: the ejection-velocity distribution $P(v_\infty)$, the ejection-time
distribution $P(t_{\rm ejec})$, and the distribution of the finite-radius ejection speed
$P(\vkick)$ measured at the first accepted ejection output (see
Section~\ref{ssec:detection}). Each subsection below isolates the dependence on
one subset of the parameter grid while holding the remaining parameters fixed,
and interprets the results in terms of the analytical framework of Section~\ref{sec:theory}.
To preserve the readability of the panels, the legends identify only the varied
parameter values; quantitative summary statistics are reported in the captions and
in the accompanying text. For any sample mean $\bar X$, we estimate the standard
error as $s_X/\sqrt{N_{\rm ejec}}$, where $s_X$ is the sample standard deviation
among successful ejections. Ejection-fraction uncertainties are binomial,
$[f_{\rm eject}(1-f_{\rm eject})/N_{\rm good}]^{1/2}$, where $N_{\rm good}$
is the number of integrations that pass the numerical-quality cuts. Fitting of
parametric models to the numerical distributions,
and quantitative comparisons with previous work, are deferred to
Section~\ref{sec:discussion}.

\begin{figure}[t]
    \centering
    \includegraphics[width=\linewidth]{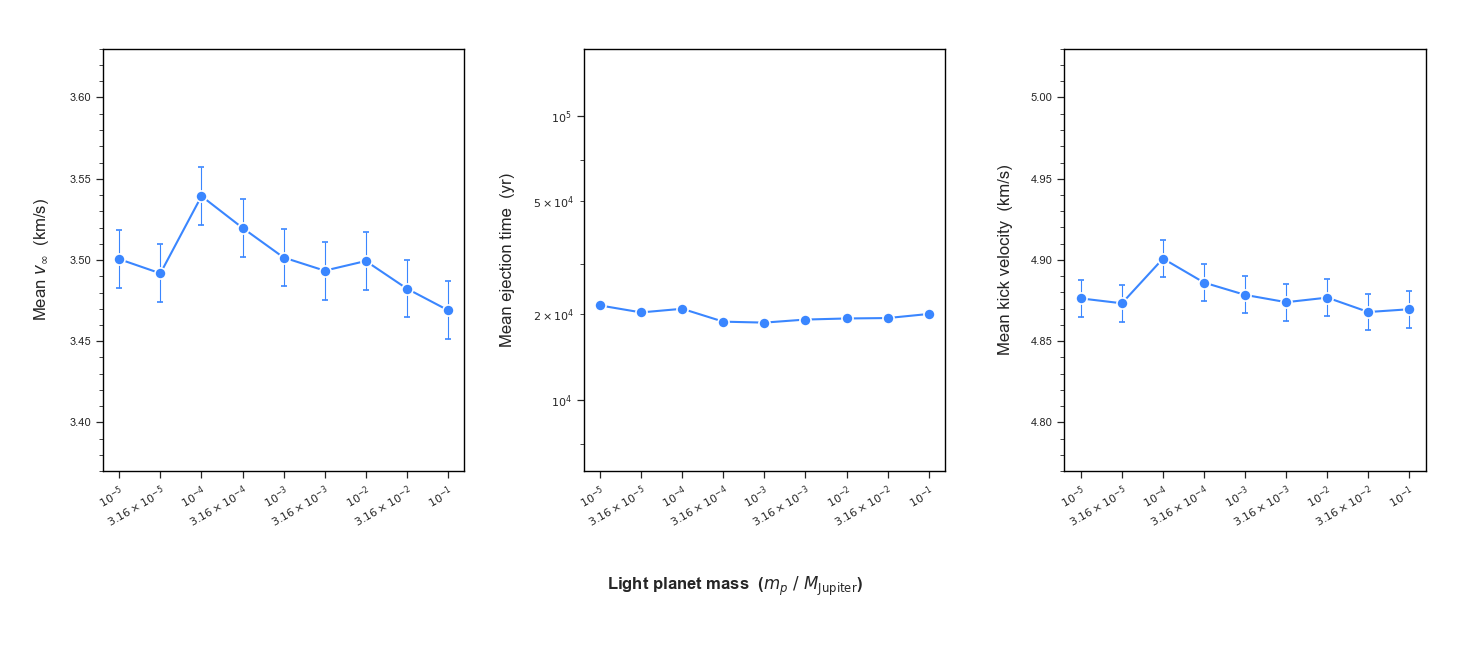}
    \caption{Mean ejection velocity $\langle v_\infty \rangle$ (left), mean ejection
    time $\langle t_{\rm ejec} \rangle$ (centre), and mean kick velocity
    $\langle \vkick \rangle$ (right) as functions of the mass of the ejected planet
    $m_p$, spanning four orders of magnitude from $10^{-5}$ to $10^{-1}\,\Mjup$. The heavy planet
    is held at $M_J = 10\,\Mjup$ and the light planet's initial semi-major axis at
    $a_p = 0.9\,a_J$. The ejection fraction rises from 96.54\% at
    $10^{-1}\,\Mjup$ to 98.21\% at $10^{-5}\,\Mjup$. All three
    statistics vary by less than the run-to-run scatter across the full mass range,
    confirming that the ejected body behaves as a test particle for
    $m_p/M_J \lesssim 10^{-2}$. Error bars on sample means are standard errors, $s_X/\sqrt{N_{\rm ejec}}$,
    where $s_X$ is the sample standard deviation of the corresponding observable; where
    smaller than the marker they are hidden.}
    \label{fig:vkick_vs_mlight}
\end{figure}
\subsection{Dependence on the mass of the ejected planet}
\label{ssec:res_mlight}

Figure~\ref{fig:vkick_vs_mlight} shows the mean ejection velocity $\langle v_\infty
\rangle$, mean ejection time $\langle t_{\rm ejec} \rangle$, and mean kick velocity
$\langle \vkick \rangle$ as a function of the mass of the ejected (light) planet,
spanning four orders of magnitude from $10^{-5}$ to $10^{-1}\,\Mjup$, with the heavy planet held
at $10\,\Mjup$ and the light planet's initial semi-major axis at $a_p = 0.9\,\ajup$.
The ejection fraction is high and nearly constant across this range: 96.54\% at
$10^{-1}\,\Mjup$, rising to 98.21\% at $10^{-5}\,\Mjup$. The three velocity and
timescale statistics show only weak, non-monotonic variation over the four orders of magnitude in
mass, with mean ejection velocities clustering around $3.4$--$3.6\,\text{km\,s}^{-1}$
and mean kick velocities around $4.8$--$5.0\,\text{km\,s}^{-1}$. Mean ejection times
scatter around $2 \times 10^{4}$\,yr with no clear trend. The near-flat response of all
three statistics to the ejected planet mass confirms that, in this regime, the dynamics
are dominated by the giant perturber: the light planet is essentially a test particle
whose own gravity does not significantly influence the encounter geometry or the energy
budget of the ejection.

This is fully consistent with the restricted three-body approximation developed in
Section~\ref{ssec:slingshot}: the encounter speed $u$ (Equation~\ref{eq:uenc}) and
the maximum ejection speed (Equation~\ref{eq:vinf_max}) both depend only on
$M_\star$, $M_J$, $a_J$, $a_p$, and $e_p$; the mass $m_p$ appears nowhere. The
slight downward trend in ejection fraction at the high-mass end ($10^{-1}\,\Mjup$,
i.e.\ Saturn-class objects) reflects the breakdown of the strict test-particle limit:
a planet massive enough to carry non-negligible angular momentum can resist being
placed into a strongly orbit-crossing configuration during secular evolution, mildly
reducing the probability of the close encounter that triggers ejection.

To quantify the residual trend more precisely, we note that the ejection fraction
decreases by $1.67$ percentage points between $10^{-5}\,\Mjup$ and $10^{-1}\,\Mjup$
(from 98.21\% to 96.54\%), while the mean ejection velocity varies by less than
$0.6\,\text{km\,s}^{-1}$ across the same range and shows no systematic monotonic
trend. The mean kick velocity is similarly stable, staying within the band
$4.8$--$5.0\,\text{km\,s}^{-1}$, and the mean ejection time fluctuates by less than
a factor of two around $2\times10^4\,\text{yr}$ with no clear mass dependence. The
total variation in all three statistics over four orders of magnitude in $m_p$ is smaller than
the run-to-run scatter within any single mass bin, confirming that the test-particle
approximation is quantitatively accurate down to the level of a few percent across
the entire range explored.

The test-particle behaviour found here is consistent with the results of
\citep{Veras2009} and \citep{Barclay2017}, both of which found that ejection
fractions and characteristic ejection speeds in multi-planet scattering are
predominantly controlled by the perturber mass rather than by the mass of the ejected
body, particularly when the mass ratio $m_p/M_J \lesssim 10^{-1}$. The slight
suppression of the ejection fraction at $10^{-1}\,\Mjup$ is also qualitatively
consistent with earlier findings that Saturn-mass objects can occasionally stabilise
inner orbits through mutual angular-momentum exchange rather than ejecting them
\citep{Chatterjee2008}.

\begin{figure}[t]
    \centering
    \includegraphics[width=\linewidth]{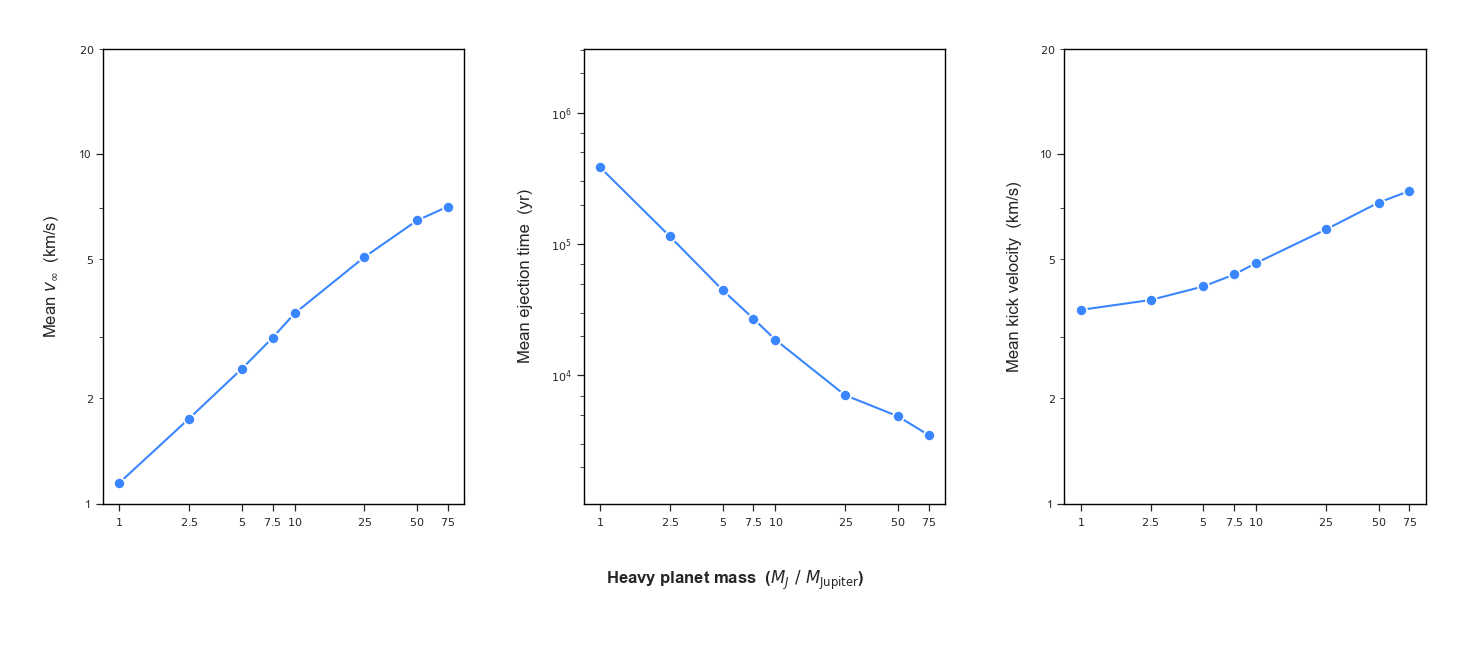}
    \caption{Mean ejection velocity $\langle v_\infty \rangle$ (left), mean ejection
    time $\langle t_{\rm ejec} \rangle$ (centre), and mean kick velocity
    $\langle \vkick \rangle$ (right) as functions of the giant-planet mass $M_J$ on a
    log-log scale. The light planet is fixed at $m_p = 10^{-3}\,\Mjup$ and
    $a_p = 0.9\,a_J$. All three statistics increase or decrease monotonically across
    $1$--$75\,\Mjup$: $\langle v_\infty \rangle$ rises from $1.15$ to
    $7.07\,\text{km\,s}^{-1}$ (power-law index $\approx 0.42$);
    $\langle t_{\rm ejec} \rangle$ falls from $3.84 \times 10^5$ to
    $3.51 \times 10^3\,\text{yr}$ (index $\approx -1.09$, consistent with the secular
    prediction $t_{\rm sec} \propto M_J^{-1}$); $\langle \vkick \rangle$ rises from
    $3.58$ to $7.84\,\text{km\,s}^{-1}$ (index $\approx 0.18$). The ejection fraction
    peaks at $\approx 98.1\%$ near $10\,\Mjup$, dips to $\approx 82.1\%$ at
    $1\,\Mjup$, and declines gradually to $\approx 95.5\%$ at $75\,\Mjup$.}
    \label{fig:vkick_vs_mheavy}
\end{figure}
\subsection{Dependence on the mass of the giant perturber}
\label{ssec:res_mheavy}

Figure~\ref{fig:vkick_vs_mheavy} shows the same three statistics as a function of the
heavy planet mass $M_J$, ranging from $1$ to $75\,\Mjup$ on a log-log scale, with the
light planet fixed at $m_p = 10^{-3}\,\Mjup$ and $a_p = 0.9\,a_J$. All three
observables vary strongly and monotonically with $M_J$. The mean ejection velocity
$\langle v_\infty \rangle$ increases from $1.15\,\text{km\,s}^{-1}$ at $1\,\Mjup$
to $7.07\,\text{km\,s}^{-1}$ at $75\,\Mjup$. The mean ejection time falls from
$3.84 \times 10^5\,\text{yr}$ at $1\,\Mjup$ to $3.51 \times 10^3\,\text{yr}$ at
$75\,\Mjup$, a decrease of nearly two orders of magnitude. The mean kick velocity
$\langle \vkick \rangle$ increases from $3.58\,\text{km\,s}^{-1}$ to
$7.84\,\text{km\,s}^{-1}$ over the same range, closely tracking $\langle v_\infty
\rangle$ but offset to higher values because $\vkick$ is measured before the planet
has fully climbed out of the residual potential well of the star--giant binary.
In all three panels the trends appear approximately linear on the log-log axes,
consistent with power-law scalings. The ejection fraction is not monotonic: it peaks at approximately $98.1\%$
near $10\,\Mjup$, dips to approximately $82.1\%$ at $1\,\Mjup$, and declines
gradually to approximately $95.5\%$ at $75\,\Mjup$.

The strong dependence on $M_J$ is expected from both the Hill-scale and slingshot
analyses. The Hill velocity scales as $\vH \propto M_J^{1/3}$
(Equation~\ref{eq:vH}), predicting a power-law growth of the median ejection speed
with that index; the slingshot upper envelope (Equation~\ref{eq:vinf_max}) grows
more steeply because $u \propto M_J^{1/3}$ enters a square root. The ejection
timescale is controlled by the rate at which the giant's secular perturbation pumps
the inner orbit into the encounter zone; from Equation~\eqref{eq:tsec},
$t_{\rm sec} \propto M_J^{-1}$, consistent with the roughly two-decade decrease in
$\langle t_{\rm ejec} \rangle$ seen over the $1$--$75\,\Mjup$ range.
The non-monotonic ejection fraction is qualitatively understandable: at $1\,\Mjup$
the Hill sphere is small, the secular timescale is long, and a significant fraction of
integrations may not complete an ejection within the allotted time; at very large
$M_J$ the Hill radius expands to the point where some inner-planet orbits are stable
within it rather than scattered out.

We can test these scalings directly from the endpoint values in
Figure~\ref{fig:vkick_vs_mheavy}. Fitting a power law $\langle O \rangle \propto
M_J^\alpha$ between $1\,\Mjup$ and $75\,\Mjup$ gives
\begin{align}
    \alpha(v_\infty) &= \frac{\ln(7.07/1.15)}{\ln 75} \approx 0.42\,, \notag \\[8pt]
    \alpha(\vkick)   &= \frac{\ln(7.84/3.58)}{\ln 75} \approx 0.18\,,  \\[8pt]
    \alpha(t_{\rm ejec}) &= \frac{\ln(3.51\times10^3/3.84\times10^5)}{\ln 75} \approx -1.09\,. \notag
\label{eq:alpha_MJ}
\end{align}

The timescale index is in excellent agreement with the secular prediction
$\alpha(t_{\rm ejec}) = -1$; the measured value of $-1.09$ differs by less than
10\%, which is well within the uncertainty of a two-point fit. The ejection-velocity
index $\alpha(v_\infty) \approx 0.42$ is steeper than the pure Hill prediction of
$1/3$: the Hill velocity at $75\,\Mjup$ relative to $1\,\Mjup$ would predict a ratio
$\langle v_\infty \rangle(75)/\langle v_\infty \rangle(1) \simeq 75^{1/3} \approx
4.2$, while the observed ratio is $7.07/1.15 \approx 6.1$. The excess is expected:
$\vH$ sets the scale for typical encounters, but the mean is pulled upward by the
high-velocity tail of the distribution, whose ceiling grows faster than $M_J^{1/3}$
through the slingshot envelope (Equation~\ref{eq:vinf_max}). The kick-velocity index
$\alpha(\vkick) \approx 0.18$ is shallower than $\alpha(v_\infty)$ because $\vkick$
is measured before the planet has fully escaped the residual potential, so the
absolute offset between $\vkick$ and $v_\infty$ compresses the dynamic range at the
low-mass end.

The non-monotonic ejection fraction admits a more quantitative interpretation in terms
of the Hill radius. At $1\,\Mjup$ the Hill radius is $\RH \approx 0.07\,a_J$
(Section~\ref{ssec:hill}), which at $a_J = 5.2\au$ corresponds to ${\sim}0.36\au$;
with $a_p = 0.9\,a_J$ the inner planet's orbit lies well outside this zone, and
ejection requires secular or resonance-overlap pumping of the eccentricity before a
close encounter becomes possible, raising the probability that the integration ends
before ejection occurs. Near $5$--$10\,\Mjup$, the overlap zone is wide enough to drive close encounters
efficiently, producing the maximum ejection fraction in the scan.
At $75\,\Mjup$ the Hill radius grows to $\RH \approx 0.29\,a_J$; now some initial
conditions place the inner planet \emph{inside} the Hill sphere, where it can be
captured into a stable resonance rather than ejected, gradually reducing the ejection
fraction again. This picture is consistent with the mild, gradual decline from
${\sim}98.1\%$ at $10\,\Mjup$ to ${\sim}95.5\%$ at $75\,\Mjup$, as opposed to the
sharper drop at $1\,\Mjup$.

\begin{figure}[t]
    \centering
    \includegraphics[width=\linewidth]{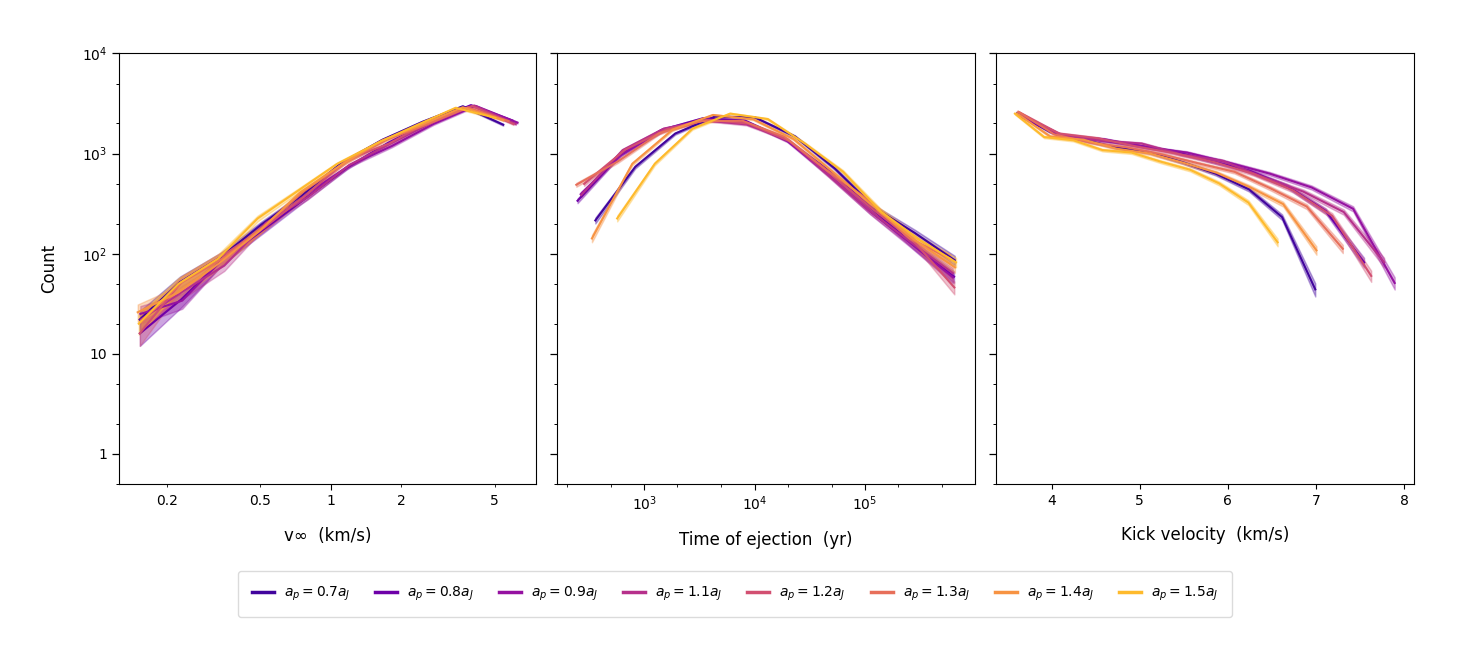}
    \caption{Distributions of ejection velocity $v_\infty$ (left), ejection time
    $t_{\rm ejec}$ (centre), and kick velocity $\vkick$ (right) for varying initial
    semi-major axis of the ejected planet $a_p$, expressed as a fraction of $a_J$
    and spanning from $0.7\,a_J$ (purple) to $1.5\,a_J$ (orange). Each coloured curve
    corresponds to a single value of $a_p/a_J$, as identified in the legend. Ejection times are
    nearly flat at $\approx 1.9 \times 10^4\,\text{yr}$ for $a_p \in [0.8, 1.2]$,
    because those configurations fall well inside the resonance-overlap zone for the
    fiducial $10\,\Mjup$ perturber ($|a_p-a_J|/a_J \lesssim 0.40$); the modest rise
    toward the two extremes reflects edge-of-zone and orbit-crossing effects rather
    than a transition controlled solely by the slower secular channel. Mean ejection velocities peak near $a_p = 0.9\,a_J$ at
    $3.50\,\text{km\,s}^{-1}$, reflecting the competition between vis-viva speed at
    orbit crossing and orbital binding energy.}
    \label{fig:sma_combined}
\end{figure}
\subsection{Dependence on the semi-major axis of the ejected planet}
\label{ssec:res_sma}

Figure~\ref{fig:sma_combined} presents the ejection velocity, ejection time, and kick
velocity distributions as functions of the initial semi-major axis $a_p$ of the light
planet, expressed as a fraction of $a_J$, for the two cases $a_p < a_J$ (inner
configurations) and $a_p > a_J$ (outer configurations) shown together. Each coloured
curve corresponds to a different value of $a_p / a_J$, ranging from $0.7\,a_J$ to
$1.5\,a_J$. The ejection fraction is high ($\gtrsim 98\%$) and approximately constant
for all values of $a_p$ shown. The shape of the
$v_\infty$ distribution is broadly similar across configurations, peaking in the range
$1$--$10\,\text{km\,s}^{-1}$, while the mean ejection and kick velocities vary only
mildly and non-monotonically, with both peaking near $a_p=0.9\,a_J$. The ejection time
distributions overlap significantly across all values of $a_p$, with means clustering
around $10^4$--$10^5\,\text{yr}$ and log-normal-like profiles on a logarithmic time
axis.

The dependence on $a_p/a_J$ is interpretable through two complementary mechanisms.
First, proximity to the giant controls how deeply a configuration lies within the
resonance-overlap zone (Equation~\ref{eq:chirikov}), and therefore how rapidly chaotic
eccentricity growth can drive the system toward a close encounter. Second, because all
runs in this scan begin on circular light-planet orbits, the eccentricity required to
reach orbit crossing depends directly on the initial separation from $a_J$. The binding
energy and encounter kinematics then determine the speed produced once crossing and a
strong encounter occur. The competition among these effects leads to the mild,
non-monotonic velocity trend seen in Figure~\ref{fig:sma_combined}.

All configurations in this scan begin with $e_p=0$, and therefore none is initially
orbit crossing. For an inner orbit with $x=a_p/a_J<1$, the eccentricity that must be
generated dynamically before crossing is possible is
\begin{equation}
    e_{\rm cross}=\frac{1}{x}-1\,,
\end{equation}
whereas for an outer orbit with $x>1$ it is
\begin{equation}
    e_{\rm cross}=1-\frac{1}{x}\,.
\end{equation}
The corresponding thresholds are $0.43$, $0.25$, and $0.11$ for $x=0.7$, $0.8$,
and $0.9$, and $0.09$, $0.17$, $0.23$, $0.29$, and $0.33$ for $x=1.1$, $1.2$,
$1.3$, $1.4$, and $1.5$, respectively. Thus the configurations at $x=0.7$ and
$x=1.5$ require substantially more eccentricity excitation than those closest to the
giant's orbit. This helps explain their modestly longer ejection times: the relevant
quantity is the eccentricity that resonance overlap or secular forcing must generate,
not a distribution of initial eccentricities.

The ejection time distributions are remarkably flat across the range $x = 0.8$--$1.2$,
with means clustering around $1.9 \times 10^4\,\text{yr}$, before rising modestly toward
both extremes ($2.45 \times 10^4\,\text{yr}$ at $x = 0.7$, $2.64 \times 10^4\,\text{yr}$
at $x = 1.5$). This flatness is not what the secular timescale prediction
(Equation~\ref{eq:tsec}, $t_{\rm sec} \propto a_p^{-3/2} \propto x^{-3/2}$) would
suggest: that scaling would predict the ejection time at $x = 0.7$ to be
$(0.7/0.8)^{-3/2} \approx 1.2$ times longer than at $x = 0.8$, while the ejection
time at $x = 1.5$ should be $(1.5/0.8)^{-3/2} \approx 0.39$ times shorter. Neither
is observed. The explanation lies in the resonance-overlap criterion
(Equation~\ref{eq:chirikov}) evaluated for the actual perturber mass used in this
scan. For $M_J = 10\,\Mjup$ around a Solar-type star,
$1.5(M_J/M_\star)^{2/7} \simeq 0.40$, so the overlap zone spans approximately
$|a_p-a_J|/a_J \lesssim 0.40$, i.e.\ $x \in [0.60,1.40]$. The configurations
$x=0.8$--$1.2$ therefore lie safely inside the chaotic overlap zone, where instability
proceeds on the fast resonance-overlap timescale rather than the slower secular one.
The two endpoint values are less clean: $x=0.7$ is still inside the nominal overlap
zone but near its inner side and has a large crossing threshold that must be generated
dynamically ($e_{\rm cross}=0.43$), while $x=1.5$ lies outside the nominal overlap
zone and relies more strongly on eccentricity pumping. Their slightly longer ejection times should
therefore be interpreted as edge/outside-of-overlap behaviour for the fiducial
$10\,\Mjup$ perturber, not as evidence that both endpoints lie outside a
Jupiter-mass overlap zone.

The mean ejection velocity shows a mild non-monotonic dependence on $x$, peaking
near $x = 0.9$ at $3.50\,\text{km\,s}^{-1}$ and declining to
$3.03\,\text{km\,s}^{-1}$ at $x = 0.7$ and $2.91\,\text{km\,s}^{-1}$ at $x = 1.5$,
a total variation of less than 20\% across the full range. This behaviour reflects
a competition between two effects that act in opposite directions. Moving the inner
orbit outward (increasing $x$) weakens its binding energy (Equation~\ref{eq:ep}),
reducing the energy increment $\Delta\epsilon$ needed for ejection and so lowering the
asymptotic speed for a given encounter strength. At the same time, the vis-viva speed
at $r = a_J$ is $v/\vJ = \sqrt{2 - 1/x}$, which rises from $0.76\,\vJ$ at $x = 0.7$
to $1.15\,\vJ$ at $x = 1.5$; outer orbits therefore arrive at the crossing radius
moving faster relative to the star, increasing the available slingshot energy through
the encounter speed $u$. The peak near $x = 0.9$ reflects the balance point where
the gain in encounter speed and the reduction in binding energy together maximise
the mean ejection velocity. The mean kick velocity $\langle \vkick \rangle$ follows
the same trend, ranging from $4.35\,\text{km\,s}^{-1}$ at $x = 0.7$ to a peak of
$4.83\,\text{km\,s}^{-1}$ at $x = 0.9$ and declining to $4.48\,\text{km\,s}^{-1}$
at $x = 1.5$, consistently offset above $\langle v_\infty \rangle$ by the residual
potential energy at the moment of ejection detection.

\begin{figure}[t]
    \centering
    \includegraphics[width=\linewidth]{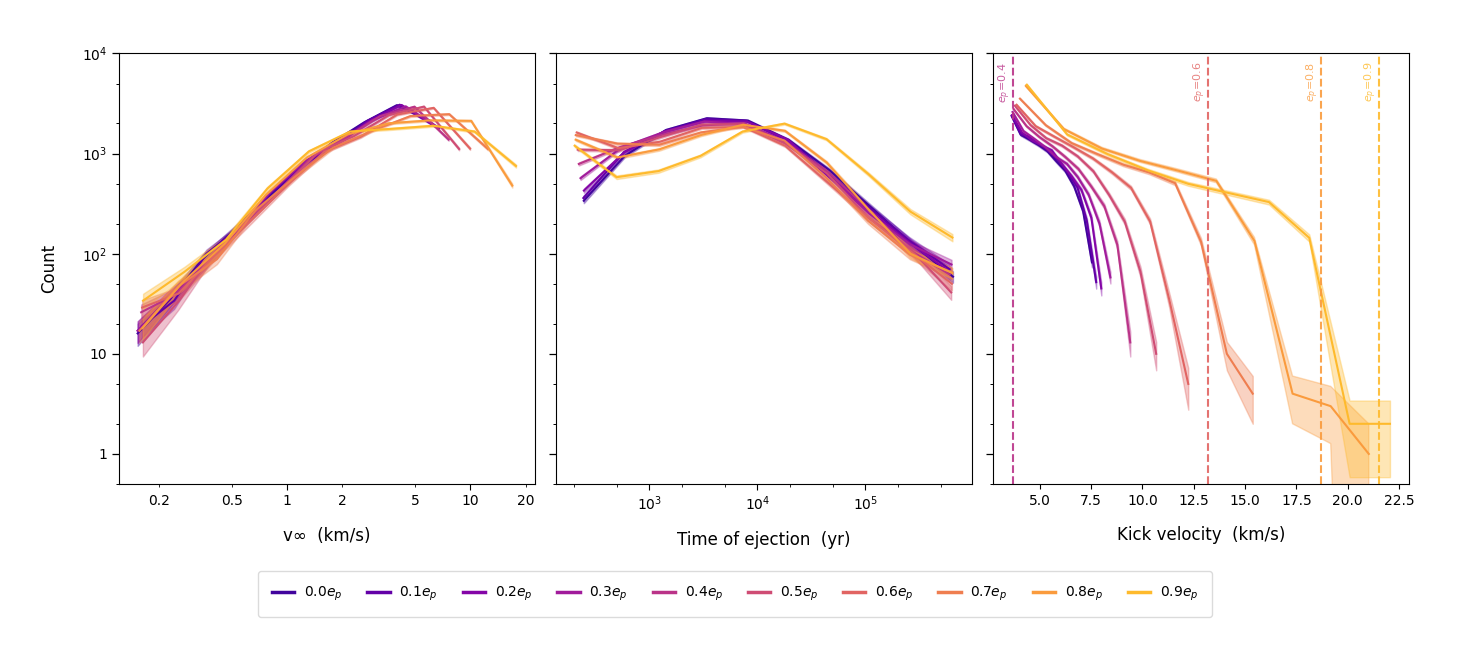}
    \caption{Distributions of ejection velocity $v_\infty$ (left), ejection time
    $t_{\rm ejec}$ (centre), and kick velocity $\vkick$ (right) as a function of the
    initial eccentricity $e_p$ of the ejected planet, for the inner configuration
    $a_p = 0.8\,a_J$. Each coloured curve corresponds to a single value of $e_p$
    ranging from 0 (purple) to 0.9 (orange) in steps of 0.1, as identified in the
    legend. Vertical dashed guide lines in the right-hand panel mark the circular-giant
    analytic upper envelope $v_{\infty,\rm max}(e_p)$ from Equation~\eqref{eq:vinf_max}
    for a sample of eccentricity bins; they are shown as an asymptotic-energy reference for the
    kick tail, not as fitted cutoffs to the finite-radius quantity $\vkick$. The $v_\infty$ distribution shifts modestly with $e_p$ (mean increases
    from $3.39$ to $5.66\,\text{km\,s}^{-1}$), while $P(\vkick)$ broadens substantially:
    the maximum $\vkick$ extends from $7.76\,\text{km\,s}^{-1}$ at $e_p = 0$ to
    $23.0\,\text{km\,s}^{-1}$ at $e_p = 0.9$, developing an extended high-velocity tail that is approximately power-law-like over
    the plotted range. Ejection fractions remain near $\approx 99\%$ throughout, with a
    slight decrease to $94.8\%$ at $e_p = 0.9$.}
    \label{fig:vkick_ep_08}
\end{figure}

\begin{figure}[t]
    \centering
    \includegraphics[width=\linewidth]{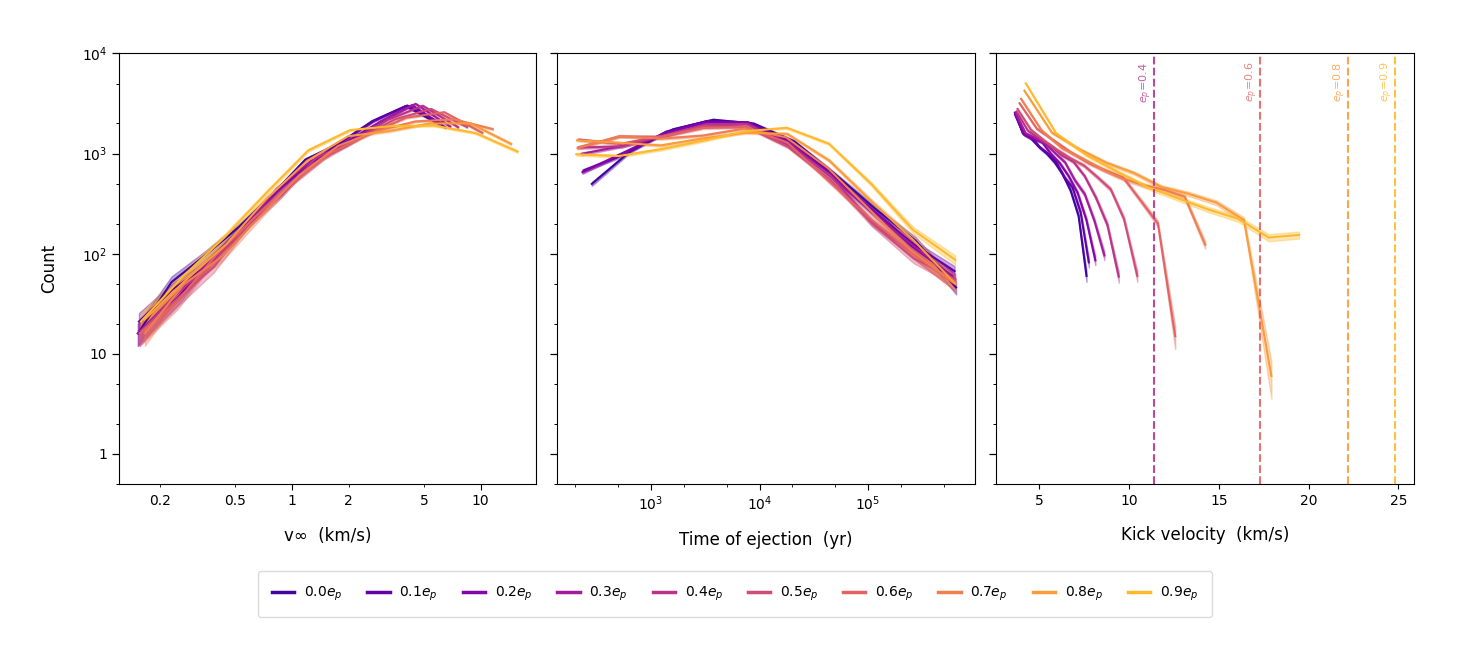}
    \caption{Same as Figure~\ref{fig:vkick_ep_08} but for the outer configuration
    $a_p = 1.2\,a_J$. The qualitative behaviour is the same: the $v_\infty$
    distribution shifts modestly (mean increases from $3.32$ to
    $5.46\,\text{km\,s}^{-1}$) while the $\vkick$ tail grows substantially (maximum
    from $7.85$ to $20.3\,\text{km\,s}^{-1}$). At every $e_p$ the outer configuration
    has a lower Tisserand parameter than the inner one
    ($\TJ = 1/x + 2\sqrt{x(1-e_p^2)}$ with $x = 1.2$ vs $0.8$), corresponding to a
    higher encounter speed $u^2/\vJ^2 = 3 - \TJ$, which produces broader velocity
    distributions throughout. Ejection fractions remain near $98$--$99\%$ across the
    full $e_p$ range. The vertical dashed guide lines in the right-hand panel have the
    same meaning as in Figure~\ref{fig:vkick_ep_08}: they show the analytic
    $v_{\infty,\rm max}(e_p)$ envelope as a reference scale for the high-velocity
    kick tail at certain eccentricities.}
    \label{fig:vkick_ep_12}
\end{figure}
\subsection{Dependence on the eccentricity of the ejected planet}
\label{ssec:res_ep}

Figures~\ref{fig:vkick_ep_08} and~\ref{fig:vkick_ep_12} show the effect of varying the
initial eccentricity $e_p$ of the light planet's orbit, for inner ($a_p = 0.8\,a_J$)
and outer ($a_p = 1.2\,a_J$) configurations respectively. In both cases $e_p$ is varied
from 0 to 0.9 in steps of 0.1. For the inner configuration (Figure~\ref{fig:vkick_ep_08})
the ejection velocity and ejection time distributions shift modestly with $e_p$: the
mean $v_\infty$ increases from roughly $3.4\,\text{km\,s}^{-1}$ at $e_p = 0$ to
$5.7\,\text{km\,s}^{-1}$ at $e_p = 0.9$, while mean ejection times decrease from
${\sim}1.9 \times 10^4\,\text{yr}$ to ${\sim}1.5 \times 10^4\,\text{yr}$. The kick
velocity distribution broadens substantially with increasing $e_p$, with the maximum
$\vkick$ extending from ${\sim}7.5\,\text{km\,s}^{-1}$ at $e_p = 0$ to
${\sim}23\,\text{km\,s}^{-1}$ at $e_p = 0.9$; the distribution develops a pronounced
high-velocity tail at large eccentricities. The outer configuration
(Figure~\ref{fig:vkick_ep_12}) exhibits qualitatively similar behaviour but with
distributions that are somewhat broader at all $e_p$, and with the high-velocity tail
of $\vkick$ extending to ${\sim}20\,\text{km\,s}^{-1}$ at $e_p = 0.9$. In both cases
the ejection fraction is high and nearly constant across the full $e_p$ range
(${\sim}98$--$99\%$ for inner and outer configurations), with a slight
decrease at the largest eccentricities in the inner configuration.

The pattern is precisely what the slingshot framework predicts. The encounter speed
$u$ (Equation~\ref{eq:uenc}) increases with $e_p$ at fixed $x$: for the inner
configuration ($x = 0.8$) it rises from $u/\vJ \approx 0$ at $e_p \to e_{\rm min}$
to $u/\vJ \approx 0.99$ at $e_p = 0.9$, while for the outer configuration ($x = 1.2$)
it reaches $u/\vJ \approx 1.10$ at the same eccentricity. Through the upper-envelope
formula (Equation~\ref{eq:vinf_max}), this translates into a rising ceiling on the
asymptotic ejection speed $v_\infty$ and therefore a corresponding reference scale for
the high-$\vkick$ tail, even though $\vkick$ itself is measured at finite radius.
The mean $v_\infty$, by contrast, grows by only a factor of ${\sim}1.67\times$ in
both configurations (from $3.39$ to $5.66\,\text{km\,s}^{-1}$ for the inner,
$3.32$ to $5.46\,\text{km\,s}^{-1}$ for the outer), while the maximum $\vkick$
grows by a factor of ${\sim}3\times$ (from $7.76$ to $23.0\,\text{km\,s}^{-1}$
for the inner, $7.85$ to $20.3\,\text{km\,s}^{-1}$ for the outer). This two-speed
response — a factor-of-two growth in the mean against a factor-of-three growth in
the tail — is the direct numerical signature of Hill-scale scattering controlling
the bulk of the distribution while the eccentricity-dependent slingshot ceiling
controls the extreme events.

A notable feature is that ejections occur even at $e_p = 0$ for both configurations,
despite the fact that a strictly circular inner orbit does not cross the giant's
orbit. Evaluating the Tisserand parameter at $e_p = 0$,
$\TJ(x{=}0.8) = 1/0.8 + 2\sqrt{0.8} \approx 3.04$ and
$\TJ(x{=}1.2) = 1/1.2 + 2\sqrt{1.2} \approx 3.02$, both slightly above the
orbit-crossing threshold $\TJ = 3$. These runs therefore reach ejection only after
secular or resonance-overlap pumping has raised $e_p$ above the crossing threshold
in situ, which is reflected in their slightly longer mean ejection times relative to
more eccentric initial conditions.

The Tisserand parameter also explains the systematic offset between the two
configurations across all $e_p$. Because $\TJ$ decreases with increasing $x$ at
fixed $e_p$, the outer configuration ($x = 1.2$) has a lower $\TJ$ than the inner
one at every eccentricity: the gap widens from $\Delta \TJ \approx 0.02$ at $e_p = 0$
to $\Delta \TJ \approx 0.24$ at $e_p = 0.9$. Via $u^2/\vJ^2 = 3 - \TJ$, the outer
configuration therefore arrives at each encounter with a higher relative speed,
producing broader distributions at all $e_p$ — consistent with the wider $P(\vkick)$
curves visible in Figure~\ref{fig:vkick_ep_12} compared to
Figure~\ref{fig:vkick_ep_08}. At $e_p = 0.9$ the outer configuration reaches
$u/\vJ \approx 1.10 > 1$, meaning the encounter velocity exceeds the giant's orbital
speed; the analytic envelope then predicts $v_{\infty,\rm max} \approx
24.8\,\text{km\,s}^{-1}$, close to the simulated maximum finite-radius kick of
${\sim}20\,\text{km\,s}^{-1}$.

Similarly, for the inner configuration at $e_p = 0.9$ the envelope gives
$v_{\infty,\rm max} \approx 21.5\,\text{km\,s}^{-1}$, close to the
simulated maximum finite-radius kick of ${\sim}23\,\text{km\,s}^{-1}$. The
slight overshoot of the simulated value relative to the analytic prediction at the
inner configuration is expected: Equation~\eqref{eq:vinf_max} is derived for coplanar
crossings at $r = a_J$, whereas the simulated ensemble includes inclined encounters
and crossings at other radii, some of which can yield marginally higher kicks.

The slight decrease in ejection fraction at the highest $e_p$ values (to $94.8\%$ at
$e_p = 0.9$ for the inner configuration) reflects the growing importance of very
close encounters. A planet on a highly eccentric orbit can pass so close to the giant
that it undergoes a strong inward deflection instead of an outward slingshot,
resulting in a tighter bound orbit rather than ejection. The deflection-angle formula
(Equation~\ref{eq:deflection}) makes this explicit: at fixed impact parameter $b$,
larger $u$ reduces $\theta$, so that at the high encounter speeds reached at
$e_p \sim 0.9$ only encounters with the smallest $b$ produce strong deflections,
and a fraction of those deflect inward. The outer configuration is less affected
($98.7\%$ at $e_p = 0.9$) because its initial semi-major axis outside $a_J$ means
it encounters the giant during infall, a geometry that statistically favours outward
rather than inward kicks.

A useful quantitative check, beyond the visual comparison in Figures~\ref{fig:vkick_ep_08}
and~\ref{fig:vkick_ep_12}, is to fit the survival function of the high-velocity tail.
For each eccentricity bin one may choose a cutoff $v_{\rm min}$, fit
$P(\vkick>v)\propto v^{-\beta}$ for $\vkick>v_{\rm min}$, and compare a log-normal-only
model against a log-normal body plus a power-law tail using an information criterion.
Here we use the emergence of the extended tail as a physical diagnostic rather than as
a fully marginalised population model; a formal tail-fit grid is deferred to
Section~\ref{ssec:limitations}.

\begin{figure}[t]
    \centering
    \includegraphics[width=\linewidth]{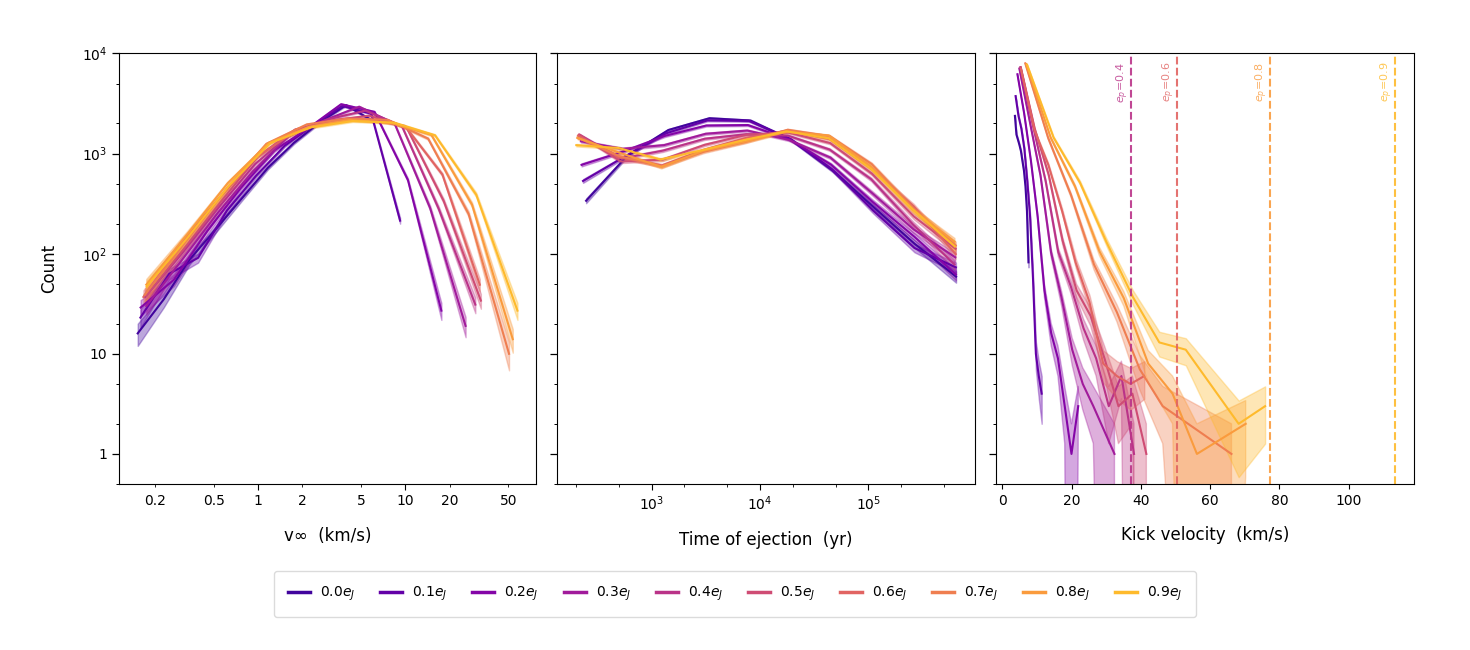}
    \caption{Distributions of ejection velocity $v_\infty$ (left), ejection time
    $t_{\rm ejec}$ (centre), and kick velocity $\vkick$ (right) as a function of the
    giant-planet eccentricity $e_J$, for the inner configuration $a_p = 0.8\,a_J$
    with $e_p = 0$. Each coloured curve corresponds to a single value of $e_J$
    ranging from 0 (purple) to 0.9 (orange) in steps of 0.1, as identified in the
    legend. The effect of $e_J$ on the tail is substantially larger than that of
    $e_p$: the maximum $\vkick$ grows from $7.76\,\text{km\,s}^{-1}$ at $e_J = 0$ to
    $79.8\,\text{km\,s}^{-1}$ at $e_J = 0.9$ (a factor of $10.3\times$, versus
    $3\times$ for the $e_p$ scan), driven by the nonlinear growth of the giant's
    pericentre speed $v_J^{\rm peri} = \vJ\sqrt{(1+e_J)/(1-e_J)}$. The mean ejection
    time decreases from $1.89 \times 10^4\,\text{yr}$ to $3.25 \times 10^3\,\text{yr}$
    as $e_J$ increases. The ejection fraction remains high throughout the scan,
    varying only at the percent level and reaching $98.7\%$ at $e_J=0.9$. Vertical
    dashed guide lines in the right-hand panel show the pericentre-enhanced analytic
    upper envelope for $v_{\infty,\rm max}$ from Equation~\eqref{eq:vinf_max_peri} for certain eccentricities;
    as in Figures~\ref{fig:vkick_ep_08}--\ref{fig:vkick_ep_12}, these are reference
    energy scales rather than fitted cutoffs to $\vkick$.}
    \label{fig:vkick_eJ_08}
\end{figure}

\begin{figure}[t]
    \centering
    \includegraphics[width=\linewidth]{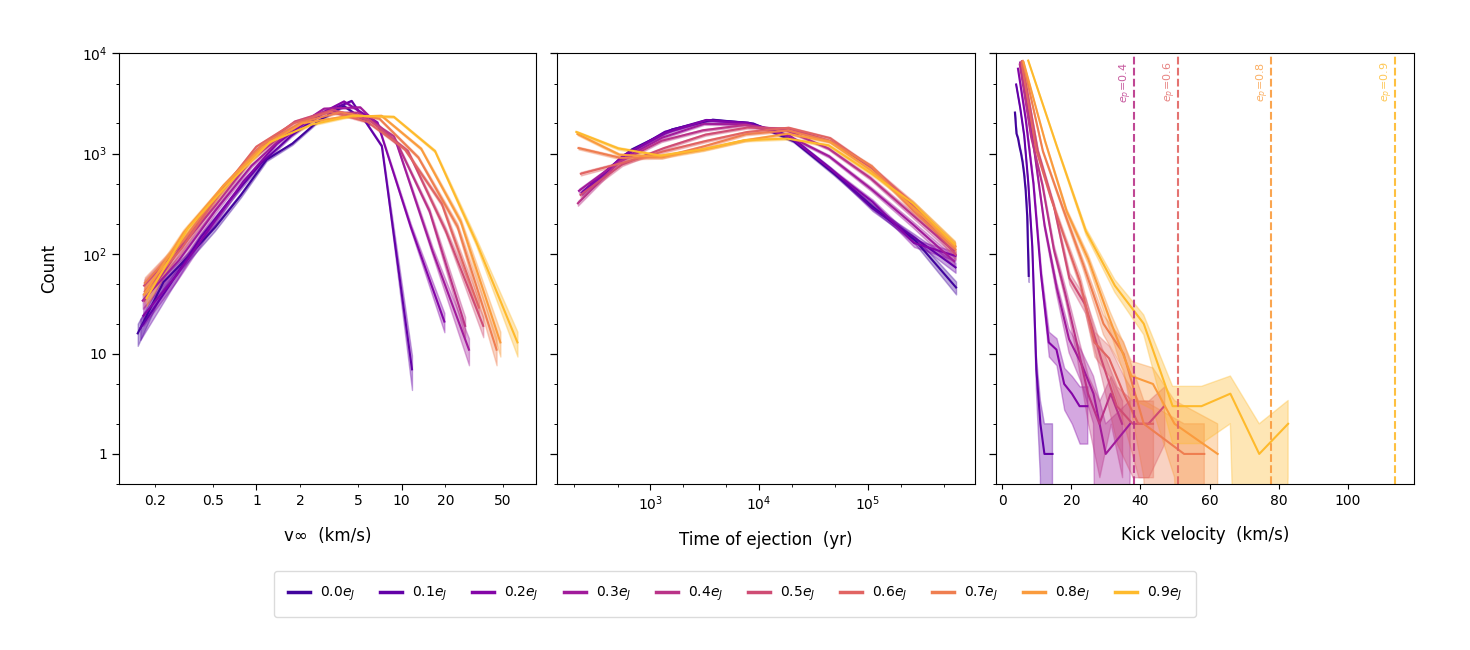}
    \caption{Same as Figure~\ref{fig:vkick_eJ_08} but for the outer configuration
    $a_p = 1.2\,a_J$. The qualitative trends mirror those of the inner configuration:
    the $\vkick$ tail extends to $86.3\,\text{km\,s}^{-1}$ at $e_J = 0.9$, and mean
    ejection times fall from $1.84 \times 10^4\,\text{yr}$ to $5.22 \times
    10^3\,\text{yr}$. The ejection fraction remains high across the scan, varying from
    about $98.7\%$ to $99.3\%$ and reaching $99.0\%$ at $e_J=0.9$; no robust
    ejection-fraction collapse is evident in the plotted values. The vertical dashed
    guide lines in the right-hand panel have the same meaning as in
    Figure~\ref{fig:vkick_eJ_08}: they show the pericentre-enhanced
    $v_{\infty,\rm max}$ envelope as a reference scale for the high-velocity kick tail at certain eccentricities.}
    \label{fig:vkick_eJ_12}
\end{figure}
\subsection{Dependence on the eccentricity of the giant perturber}
\label{ssec:res_eJ}

Figures~\ref{fig:vkick_eJ_08} and~\ref{fig:vkick_eJ_12} show the corresponding
results for variation of the giant planet's orbital eccentricity $e_J$, again for
inner ($a_p = 0.8\,a_J$) and outer ($a_p = 1.2\,a_J$) configurations. The effect
of $e_J$ on all three distributions is substantially larger than that of $e_p$.
For the inner configuration (Figure~\ref{fig:vkick_eJ_08}), the mean ejection
velocity increases overall from $3.39\,\text{km\,s}^{-1}$ at $e_J=0$ to
$6.97\,\text{km\,s}^{-1}$ at $e_J=0.9$, with the largest changes occurring toward
the high-eccentricity end of the scan. More striking is the behaviour of the kick velocity: the maximum
$\vkick$ recorded grows from ${\sim}7.8\,\text{km\,s}^{-1}$ at $e_J = 0$ to
${\sim}79.7\,\text{km\,s}^{-1}$ at $e_J = 0.9$, and the distribution at high $e_J$
develops an extended, approximately power-law-like high-velocity tail that is absent at low eccentricity. Mean ejection
times decrease steadily with $e_J$, from ${\sim}1.9 \times 10^4\,\text{yr}$ at
$e_J = 0$ to ${\sim}3.2 \times 10^3\,\text{yr}$ at $e_J = 0.9$. The outer
configuration (Figure~\ref{fig:vkick_eJ_12}) shows the same qualitative trend,
with the maximum kick velocity reaching ${\sim}86.3\,\text{km\,s}^{-1}$ at
$e_J = 0.9$. Ejection fractions remain high in both panels across the full
$e_J$ range, at roughly $98.5$--$99.3\%$ in the plotted values. Thus increasing
$e_J$ mainly redistributes the velocity tail and shortens the ejection time; it does
not appreciably suppress the probability of ejection within the integration time.

The substantially stronger effect of $e_J$ compared to $e_p$ has a natural explanation.
The giant's eccentricity directly modulates the encounter speed through two channels
that both act in the same direction. First, a giant on an eccentric orbit periodically
approaches the inner planet more closely than a circular orbit of the same semi-major
axis would; at pericentre $q_J = a_J(1-e_J)$, the giant sweeps through the inner
system faster and with a larger radial velocity component, increasing the velocity
mismatch with the inner planet. Second, and more importantly, the encounter velocity
formula (Equation~\ref{eq:uenc}) was derived for a giant on a circular orbit; when
$e_J > 0$, the giant's own instantaneous speed differs from $\vJ$ by of order
$e_J \vJ$, which modifies the relative velocity $\bm{u} = \bm{v}_p - \bm{v}_J(t)$
at each point along the orbit. At high $e_J$ the giant can pass through its own
pericentre while the inner planet is near the crossing radius, producing encounter
speeds much larger than Equation~\eqref{eq:uenc} predicts for the circular case.
A first-order circular-orbit estimate can be obtained by treating the giant's
velocity perturbation as $\Delta v_J\sim e_J\vJ$:
\begin{equation}
    \frac{v_{\infty,\rm max}^2}{\vJ^2}
    \sim
    4\,\frac{u_{\rm lin}}{\vJ} - \frac{1}{x}\,,
    \qquad
    u_{\rm lin} \sim u(e_p) + e_J\vJ\,.
\label{eq:vinf_max_eJ}
\end{equation}
This expression is useful as a conservative scaling, but it underestimates the most
extreme kicks because the fastest encounters with an eccentric giant occur near
pericentre. The giant's instantaneous pericentre speed is
\begin{equation}
    v_J^{\rm peri} = \vJ\sqrt{\frac{1+e_J}{1-e_J}}\,,
\label{eq:vJperi}
\end{equation}
which grows nonlinearly as $e_J$ approaches unity. For a Jupiter analog
($\vJ\approx13.1\,\text{km\,s}^{-1}$), $v_J^{\rm peri}$ reaches $17.9$, $22.7$,
$31.2$, and $57.1\,\text{km\,s}^{-1}$ at $e_J=0.3$, $0.5$, $0.7$, and $0.9$,
respectively. The appropriate pericentre-enhanced envelope is therefore
\begin{equation}
    v_{\infty,\rm max}^2
    \sim 4\,v_J^{\rm peri}\,u_{\rm eff} - \frac{\vJ^2}{x}\,,
\label{eq:vinf_max_peri}
\end{equation}
where $u_{\rm eff}$ is a geometry-dependent encounter speed. Equation~\eqref{eq:vinf_max_eJ}
with $u_{\rm lin}\simeq e_J\vJ$ gives only $v_{\infty,\rm max}\sim20\,\text{km\,s}^{-1}$
for $e_J=0.9$ and $x=0.8$. By contrast, Equation~\eqref{eq:vinf_max_peri} shows that
the simulated value $v_{\infty,\rm max}\simeq79.7\,\text{km\,s}^{-1}$ corresponds to
$u_{\rm eff}\simeq29\,\text{km\,s}^{-1}\simeq0.5\,v_J^{\rm peri}$, a favourable but
not maximal pericentre slingshot. The strict choice $u_{\rm eff}\sim v_J^{\rm peri}$
would give an envelope near $110\,\text{km\,s}^{-1}$, so the measured $80$--$86\,\text{km\,s}^{-1}$
maxima lie comfortably below the pericentre-corrected ceiling. This contrasts sharply
with the $e_p$ case, where the encounter speed from Equation~\eqref{eq:uenc} is bounded
by $\sqrt{3}\,\vJ$ even at $e_p=1$.

A direct side-by-side comparison with the $e_p$ scan (Section~\ref{ssec:res_ep})
quantifies the difference. For the inner configuration ($x = 0.8$), varying $e_J$
from 0 to 0.9 grows the mean $v_\infty$ by a factor of $2.06\times$ (from $3.39$ to
$6.97\,\text{km\,s}^{-1}$) and the maximum $\vkick$ by $10.3\times$ (from $7.76$ to
$79.7\,\text{km\,s}^{-1}$). The corresponding factors for the $e_p$ scan over the
same range are $1.67\times$ and $3.0\times$. The $e_J$ effect on the tail is
therefore ${\sim}3.5\times$ larger, while its effect on the mean ejection speed is
only ${\sim}1.2\times$ larger — confirming that both parameters primarily act on the
tail, but through mechanisms of very different strength.

The mean ejection time is also more strongly suppressed by $e_J$ than by $e_p$.
The orbit-averaged perturbation from an eccentric giant is enhanced by a factor of
$(1-e_J^2)^{-3/2}$ relative to the circular case, which shortens the secular
timescale by the same factor (from $e_J = 0.8$: predicted ratio $0.22$, measured
$0.52$; the measured decrease is shallower because at moderate $e_J$ the impulsive
channel begins to compete with secular pumping before the latter can complete).
From $e_J = 0$ to $e_J = 0.8$, the inner-configuration mean ejection time decreases
by a factor of ${\sim}1.9\times$, compared with only ${\sim}1.2\times$ over the
equivalent $e_p$ range.

The ejection fraction is therefore not the anomalous quantity in the $e_J$ scan;
the robust effect is the growth of the high-velocity tail. This separation is physically
useful. It indicates that eccentric giant orbits do not primarily make ejection more or
less likely in the present grid; rather, they change the energy scale of the subset of
encounters that already lead to escape. Any apparent single-bin suppression in an
intermediate analysis should therefore be checked against the final post-quality-cut
ejection counts before being interpreted as a resonance effect.

A formal tail analysis can be performed using the same survival-function procedure
described above. The relevant question is not merely whether the high-velocity histogram
looks straight on log--log axes, but whether a log-normal body plus an approximately
power-law tail is statistically preferred over a single-component model. A practical
criterion would require at least tens of events above the fitted cutoff, an acceptable
Kolmogorov--Smirnov probability for the tail fit, and a substantial information-criterion
preference for the two-component model. If future reruns reveal a genuine ejection-fraction
dip in any particular $e_J$ bin, it should be tested with additional random phases and by
recording the resonant angle histories.

\section{Implications for the FFP Population}
\label{sec:massfunc}

The numerical results of Section~\ref{sec:results} characterise the ejection output
of individual three-body systems as functions of the system architecture. We now
connect these results to three observationally relevant quantities: the present-day
Galactic FFP number density, the FFP velocity dispersion, and the FFP mass function.
In each case we derive what can be obtained analytically or semi-analytically from
our results and from independent observational constraints, and flag the remaining
purely numerical steps for future work.

\subsection{Translating ejection yields into a Galactic FFP population}
\label{ssec:galactic_model}

The present-day number density of FFPs per unit volume in the Galactic disc can
be written as
\begin{equation}
    n_{\rm FFP} = n_\star \int_0^{t_{\rm now}}
    \dot{n}_{\rm eject}(t)\,{\rm d}t
    \;\approx\;
    n_\star \, f_{\rm sys} \, \langle N_{\rm eject} \rangle \, f_{\rm eject}\,,
\label{eq:nFFP}
\end{equation}
where $n_\star$ is the local stellar number density, $f_{\rm sys}$ is the fraction of
stars hosting a suitable giant-plus-inner-planet architecture, $\langle N_{\rm eject}
\rangle$ is the mean number of planets ejected per unstable system, and $f_{\rm eject}$
is the ejection fraction per system, which we measure to be
$f_{\rm eject} \approx 0.95$--$0.98$ across nearly the entire parameter grid
(Sections~\ref{ssec:res_mlight}--\ref{ssec:res_eJ}).

For the remaining factors we adopt observational estimates from radial-velocity and
transit surveys. The fraction of FGK stars hosting a giant planet with
$M_J \in [0.3, 10]\,\Mjup$ at $a_J \in [1, 20]\au$ is
$f_{\rm giant} \approx 10\%$ \citep{Cumming2008}, and roughly half of those systems
host at least one inner companion capable of being ejected, giving
$f_{\rm sys} \approx f_{\rm giant} \times 0.5 \approx 5\%$.
With $\langle N_{\rm eject} \rangle \approx 1$ per unstable system and
$n_\star \approx 0.1\,\text{pc}^{-3}$ in the Solar neighbourhood, we obtain an
order-of-magnitude estimate
\begin{equation}
    n_{\rm FFP} \approx 0.1\,\text{pc}^{-3}
    \times 0.05
    \times 1
    \times 0.97
    \approx 5 \times 10^{-3}\,\text{pc}^{-3}\,.
\label{eq:nFFP_numerical}
\end{equation}
This estimate should be interpreted as an order-of-magnitude contribution from one
specific channel rather than as a complete census of FFPs. It counts only hierarchical
three-body ejections and assumes one ejected planet per unstable system. If the unstable
fraction of giant-bearing systems is higher, or if multiple inner planets are ejected per
system, $n_{\rm FFP}$ scales proportionally. Conversely, if only a subset of giant-bearing
systems enters the unstable architecture considered here, the normalisation is reduced.
The comparison with microlensing-inferred FFP abundances is therefore best made after
convolving this channel with a population model for giant-planet occurrence, eccentricity,
and system multiplicity.

\subsection{Predicted FFP velocity dispersion}
\label{ssec:veldisp}

After the planet has escaped the host system, its Galactic velocity is the vector sum of
the host star's Galactic velocity and the asymptotic relative velocity $v_\infty$.
Assuming that the ejection direction is uncorrelated with the stellar Galactic velocity,
the three-dimensional velocity dispersion of the FFP population is
\begin{equation}
    \sigma_{\rm FFP}^2 = \sigma_\star^2 + \langle v_\infty^2 \rangle\,,
\label{eq:sigmaFFP}
\end{equation}
where the average must ultimately be taken over both the simulated ejection-velocity
distributions and the occurrence distribution of planetary-system architectures. The
finite-radius quantity $\vkick$ is not appropriate in Equation~\eqref{eq:sigmaFFP},
because part of it is lost as the planet climbs out of the residual stellar potential
(Section~\ref{ssec:vesc}).

For the thin disc, $(\sigma_U,\sigma_V,\sigma_W)\approx(35,25,20)\,\text{km\,s}^{-1}$
\citep{Holmberg2009}, corresponding to $\sigma_\star\approx47.4\,\text{km\,s}^{-1}$.
Across most of the present grid, characteristic values of $v_\infty$ are only a few
$\text{km\,s}^{-1}$. A representative rms asymptotic speed of
$4$--$7\,\text{km\,s}^{-1}$ would give
\begin{equation}
    \sigma_{\rm FFP}\approx47.6\text{--}47.9\,\text{km\,s}^{-1}\,,
\end{equation}
a change of at most approximately one percent relative to the stellar baseline. The
bulk of the ejected population is therefore expected to be kinematically difficult to
distinguish from its parent stellar population through a dispersion measurement alone.
A precise value requires a population prior for $M_J$, $a_J$, $e_J$, $a_p$, and the
other initial conditions; the controlled parameter grid used here is not such a prior.

The eccentric-giant simulations do demonstrate that rare, much faster ejections are
possible. However, the maximum velocity in a finite Monte Carlo sample cannot be used
to infer the fraction of the population in this tail. For a chosen asymptotic threshold
$v_0$, the relevant quantity is the event-level exceedance fraction
\begin{equation}
    F_{>v_0}(\boldsymbol{\Theta})=
    \frac{N[v_\infty>v_0\mid\boldsymbol{\Theta}]}{N_{\rm ejec}(\boldsymbol{\Theta})}\,,
\end{equation}
where $\boldsymbol{\Theta}$ denotes a simulated system configuration. The Galactic
fraction above the threshold is then
\begin{equation}
    f_{>v_0}=\int {\rm d}\boldsymbol{\Theta}\,
    p(\boldsymbol{\Theta})F_{>v_0}(\boldsymbol{\Theta})\,.
\label{eq:tail_fraction}
\end{equation}
For the inner-configuration eccentric-giant scan shown in Figure~\ref{fig:vkick_eJ_08}, the event-level exceedance fraction $F_{>v_0}(\boldsymbol{\Theta})$ increases with the giant's eccentricity. It vanishes at $e_J = 0$, and  $F_{>20\,\text{km\,s}^{-1}}(\boldsymbol{\Theta})$ rises to 5.87\% at $e_J = 0.9$, while $F_{>50\,\text{km\,s}^{-1}}(\boldsymbol{\Theta})$ stays below 0.2\% across the full range and only becomes nonzero for $e_J \ge 0.7$. The outer configuration scan shown in Figure~\ref{fig:vkick_eJ_12} reveals a similar but weaker trajectory. At $e_J = 0.9$, $F_{>20\,\text{km\,s}^{-1}}(\boldsymbol{\Theta})$ reaches 2.48\%, and $F_{>50\,\text{km\,s}^{-1}}(\boldsymbol{\Theta})$ equals 0.11\%. These trends confirm that the high-velocity tail is dominated by the most eccentric configurations, and that at the highest $e_J$ values the inner configuration produces larger exceedance fractions.

Thus the simulations establish the existence of a potentially observable high-velocity
tail, but its abundance cannot be quoted until the full $v_\infty$ samples are combined
with an empirically motivated architecture distribution.


\subsection{Predicted mass function and microlensing signature}
\label{ssec:mf}

The mass function of the present-day Galactic FFP population is set by the
convolution of the planet occurrence rate ${\rm d}N_{\rm planet}/{\rm d}m_p$ with
the ejection probability as a function of $m_p$. The key result of
Section~\ref{ssec:res_mlight} is that the ejection fraction is nearly independent of
$m_p$ over four orders of magnitude, varying by only $2.3$ percentage points between
$10^{-5}\,\Mjup$ and $10^{-1}\,\Mjup$. The FFP mass function therefore directly
traces the underlying planet occurrence rate:
\begin{equation}
    \frac{{\rm d}N_{\rm FFP}}{{\rm d}m_p}
    \approx f_{\rm eject}(m_p)\,\frac{{\rm d}N_{\rm planet}}{{\rm d}m_p}\,,
    \qquad
    f_{\rm eject}(m_p) \simeq \bar f_{\rm eject} \approx 0.97
    \quad\text{for}\quad 10^{-5}\leq \frac{m_p}{\Mjup}\leq10^{-1}\,.
\label{eq:massfunc}
\end{equation}
The residual variation across the simulated interval is comparable to the small
configuration-to-configuration scatter and does not justify a mass-dependent fit. To the
accuracy supported by the present scan,
${\rm d}N_{\rm FFP}/{\rm d}m_p \approx 0.97\,{\rm d}N_{\rm planet}/{\rm d}m_p$.

The planet occurrence rate at sub-Earth masses is not directly constrained by current
observations, but \textit{Kepler} data suggest that ${\rm d}N/{\rm d}(\log m_p)$
is approximately flat or rising toward lower masses in the sub-Neptune regime
\citep{Howard2012}. Extrapolating this to sub-Earth masses, the FFP mass function
would also be flat or rising in $\log m_p$, making the sub-Earth population
numerically dominant, consistent with the ultrashort events reported by
\citep{Mroz2020}.

\begin{figure}[t]
\centering
\includegraphics[width=0.85\linewidth]{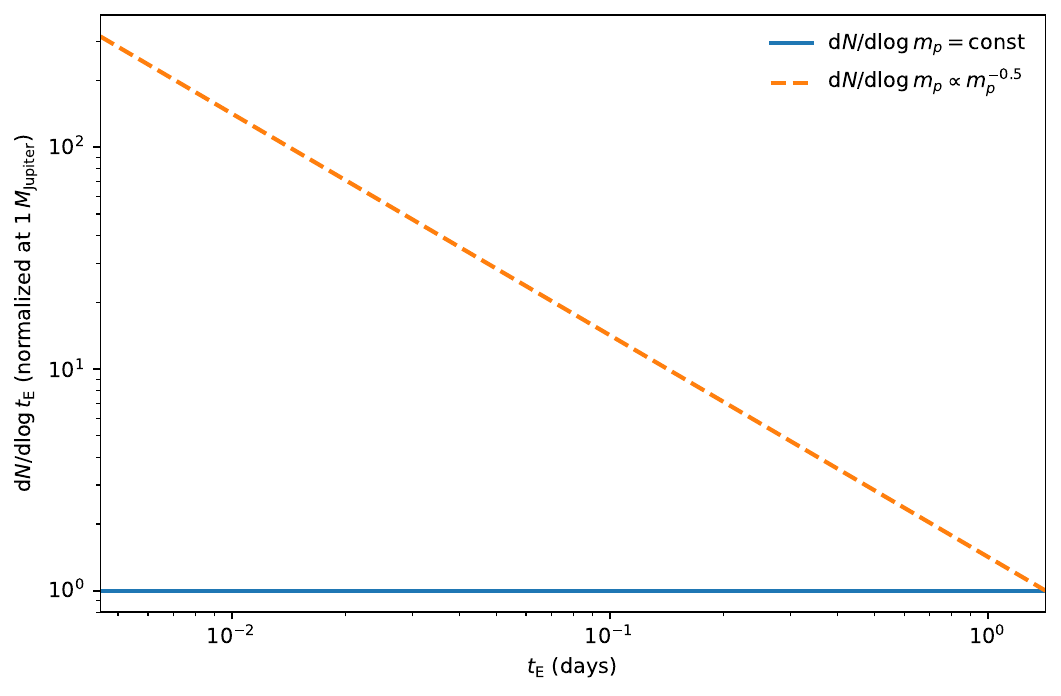}
\caption{Transformed lens-timescale distribution ${\rm d}N/{\rm d}(\log t_{\rm E})$
implied by two illustrative planet mass functions. The solid curve is flat in
$\log m_p$, ${\rm d}N/{\rm d}\log m_p=\text{const}$, and the dashed curve rises toward
lower masses as $m_p^{-0.5}$; both are normalised to unity at one Jupiter mass. The
conversion from $m_p$ to $t_{\rm E}$ uses Equation~\eqref{eq:tE} with
$D_L=4\,\text{kpc}$, $D_S=8\,\text{kpc}$, and
$\mu_{\rm rel}=8\,\text{mas\,yr}^{-1}$. The nearly constant ejection efficiency in
Equation~\eqref{eq:massfunc} changes the normalisation but not the shape. This is not
yet a microlensing event-rate prediction: lensing cross-section, Galactic phase-space
weights, cadence, blending, and detection efficiency have not been applied.}
\label{fig:tE_dist}
\end{figure}

The observable proxy for $m_p$ in microlensing is the Einstein-ring crossing time
$t_{\rm E}$. For a lens at $D_L = 4\,\text{kpc}$ toward a source at $D_S = 8\,\text{kpc}$,
\begin{equation}
    \theta_{\rm E} = 0.031\,\text{mas}
    \left(\frac{m_p}{\Mjup}\right)^{1/2},
    \qquad
    t_{\rm E} \approx 1.42\,\text{day}
    \left(\frac{m_p}{\Mjup}\right)^{1/2}
    \left(\frac{\mu_{\rm rel}}{8\,\text{mas\,yr}^{-1}}\right)^{-1}.
\label{eq:tE}
\end{equation}
Figure~\ref{fig:tE_dist} shows the mass function transformed into ${\rm d}N/{\rm d}(\log t_{\rm E})$
for two representative occurrence-rate assumptions: flat in $\log m_p$ (solid) and
rising as ${\rm d}N/{\rm d}(\log m_p) \propto m_p^{-0.5}$ (dashed). The flat case
produces a distribution that is uniform across all $t_{\rm E}$ bins — equal numbers
of lenses per decade in crossing time before lensing and survey weights are applied — whereas the rising case predicts
substantially more lenses at $t_{\rm E} \lesssim 0.1\,\text{day}$ (sub-Earth masses),
with the two models diverging by more than two orders of magnitude at the shortest
timescales shown. This divergence is a potentially discriminating prediction for the
\textit{Roman} Wide Field Survey \citep{Penny2019}, but a true event-rate prediction
requires multiplication by the microlensing cross-section, the Galactic lens--source
phase-space distribution, and the survey detection efficiency.

\begin{figure}[t]
\centering
\includegraphics[width=0.85\linewidth]{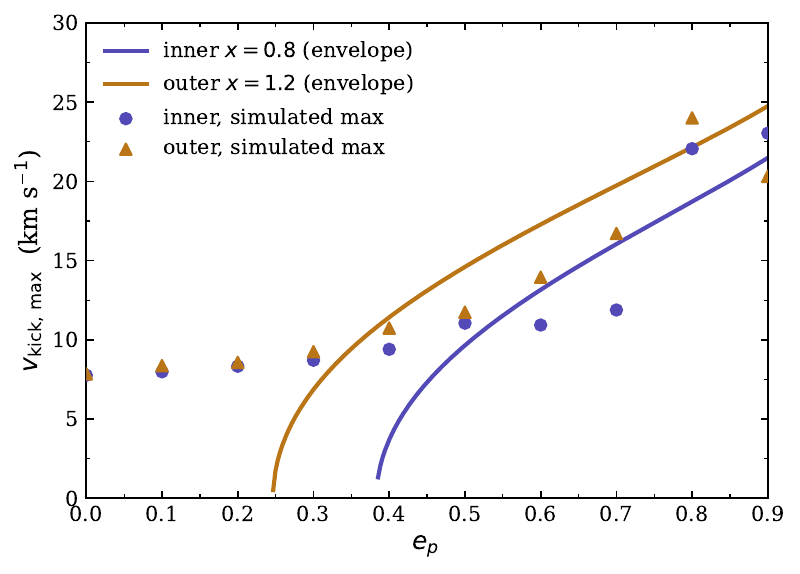}
\caption{Analytic upper envelope of the ejection speed $v_{\infty,\rm max}$
(Equation~\ref{eq:vinf_max}, solid curves) as a function of the initial eccentricity
$e_p$ of the ejected planet, for the inner ($x = a_p/a_J = 0.8$, purple) and outer
($x = 1.2$, amber) configurations with a Jupiter-analog perturber on a circular orbit
($e_J = 0$, $\vJ = 13.1\,\text{km\,s}^{-1}$). Circles and triangles show the
measured maximum $\vkick$ values from Figures~\ref{fig:vkick_ep_08}
and~\ref{fig:vkick_ep_12} respectively. Because $\vkick$ is measured at finite radius,
this comparison is approximate; nevertheless, the envelope correctly tracks the ceiling
of the simulated high-velocity tail at $e_p \gtrsim 0.5$; the excess of the simulated
points at lower $e_p$ reflects ejections occurring after secular pumping of the
eccentricity to values above the initial $e_p$.}
\label{fig:envelope}
\end{figure}

The analytic upper-envelope formula (Equation~\ref{eq:vinf_max}) provides a further
connection between our simulations and the microlensing signal.
Figure~\ref{fig:envelope} shows the predicted $v_{\infty,\rm max}$ as a function of
$e_p$ for the inner and outer configurations, together with the measured simulation
maxima from Figures~\ref{fig:vkick_ep_08} and~\ref{fig:vkick_ep_12}. The envelope
traces the simulated maxima closely at $e_p \gtrsim 0.5$, confirming that the
slingshot framework (Section~\ref{ssec:slingshot}) correctly predicts the ceiling of
the kick-velocity distribution. At lower $e_p$ the simulated values lie above the
coplanar circular-giant formula because those runs reach ejection only after secular
eccentricity pumping, as discussed in Section~\ref{ssec:res_ep}. The envelope can
therefore be used as a prior on the maximum kick velocity when modelling the
microlensing parallax signal of individual FFP events: a measured $v_\infty$ above
the envelope for the inferred $e_p$ would indicate either a highly eccentric giant
perturber or a non-coplanar encounter geometry.

The asymptotic ejection velocity adds a contribution to $\mu_{\rm rel}$ beyond the
typical stellar proper motion. For a representative $v_\infty \approx 5\,\text{km\,s}^{-1}$ at
$D_L = 4\,\text{kpc}$, the additional proper-motion contribution is
$\delta\mu \approx 0.26\,\text{mas\,yr}^{-1}$, a ${\sim}3\%$ correction to
$\mu_{\rm rel}$ that is negligible for population-level predictions but may be
measurable in individual events with well-constrained microlensing parallaxes.

\section{Discussion}
\label{sec:discussion}

\subsection{Comparison with previous numerical studies}
\label{ssec:comparison}

The most directly comparable prior work on kick velocities from planet--planet
scattering is \citep{Veras2009}, who studied ejection outcomes in two-planet systems
around Solar-type hosts and found that ejection fractions are high
($\gtrsim 90\%$ for mass ratios $m_p/M_J \lesssim 0.1$) and that the ejected body
behaves as a test particle once this mass ratio is reached. Our results confirm and
quantify this: the ejection fraction varies by only $2.3$ percentage points across
four orders of magnitude in $m_p$ (Section~\ref{ssec:res_mlight}), and all three ejection
statistics are insensitive to $m_p$ at the level of the run-to-run scatter. The
test-particle limit is thus well established by $m_p/M_J \lesssim 10^{-2}$ in the
hierarchical three-body geometry studied here.

\citep{Chatterjee2008} and \citep{Raymond2010} studied planet--planet scattering in
systems with two to five planets, finding that the eccentricity distribution of the
surviving planet and the ejection velocity of the escapee are both strongly controlled
by the mass of the most massive body in the system. This is consistent with our
$M_J$-dependence results (Section~\ref{ssec:res_mheavy}): the mean ejection velocity
grows as $M_J^{0.42}$ and the ejection timescale falls as $M_J^{-1.09}$, with the
latter in excellent agreement with the secular timescale scaling $t_{\rm sec} \propto
M_J^{-1}$ (Equation~\ref{eq:tsec}). Our configuration, however, is more controlled
than those multi-planet studies: by fixing the outer giant and varying only one inner
planet at a time we isolate the dependence on each parameter without the
confounding influence of additional bodies.

\citep{Barclay2017} performed a focused study of Earth-mass planet ejection from
systems with a single giant perturber, finding characteristic ejection speeds of a
few km\,s$^{-1}$ and ejection fractions near $\sim90\%$ for tight inner orbits.
Our results are broadly consistent: for the fiducial $10\,\Mjup$ perturber at
$x = 0.9$ we find $\langle v_\infty \rangle \approx 3.5\,\text{km\,s}^{-1}$ and
ejection fractions of $97$--$99\%$, with the somewhat higher fractions attributable
to our wider range of initial eccentricities and our use of the IAS15 integrator
rather than a mixed-variable symplectic scheme, which handles close encounters more
accurately. The key advance of our work over \citep{Barclay2017} is the controlled mapping of the
kick-velocity distribution over the principal dynamical parameters in a hierarchical
three-body setup and the identification of the upper-envelope formula
(Equation~\ref{eq:vinf_max}) as a predictor of the tail, which was not characterised in
that study.

\citep{Rasio1996} established the gravitational-slingshot picture for planet--planet
scattering and showed that the ejection velocity scales with the orbital speed of the
perturber. Our Equations~\eqref{eq:uenc} and~\eqref{eq:vinf_max} are a quantitative
development of that picture, connecting the kick-velocity ceiling explicitly to the
eccentricity through the Tisserand parameter (Section~\ref{ssec:tisserand}).
The two-speed response we identify — mean velocity controlled by Hill-scale
scattering, tail controlled by the slingshot envelope — provides a cleaner
decomposition of the velocity distribution than was available from the earlier
analytical work.

\subsection{Sensitivity to initial condition choices}
\label{ssec:sensitivity}

Several choices in our numerical setup deserve scrutiny.

\paragraph{Orbital phase and argument of pericentre.}
The initial mean anomalies are randomised uniformly for each run, and the inclinations
are drawn uniformly from $0^\circ$--$3^\circ$ as described in Section~\ref{ssec:grid}.
The arguments of pericentre, however, are not independently randomised in the present
runs; they are left at their \textsc{rebound} defaults. The results reported in
Section~\ref{sec:results} are therefore averaged over initial orbital phase and small
inclination differences, but not over the full apsidal-orientation phase space at fixed
$(M_J, a_J, m_p, a_p, e_p, e_J, i)$. The scatter within each parameter bin — visible
as the point-to-point noise in Figures~\ref{fig:vkick_vs_mlight} and~\ref{fig:vkick_vs_mheavy}
— reflects the residual phase dependence after this partial marginalisation. For the mean
statistics this scatter is small (sub-percent for the ejection fraction, $\lesssim 10\%$
for the mean velocity), but the maximum kick velocity in any given bin is sensitive to
the fraction of runs that happen to align for a near-optimal slingshot encounter. The
upper envelope formula (Equation~\ref{eq:vinf_max}) should therefore be interpreted as
a ceiling over the sampled ensemble, not as a prediction for any individual run.

\paragraph{Eccentricity prior.}
In the eccentricity scans presented here, both the inner-planet eccentricity and the
giant-planet eccentricity are sampled uniformly over $[0,0.9]$ in steps of $0.1$.
This uniform scan is a controlled numerical grid rather than an occurrence-rate prior.
The observed RV eccentricity distribution for giant planets is approximately thermal
($P(e) \propto e$ or $2e$) for $e \gtrsim 0.2$ \citep{Juric2008}, which weights
more heavily toward high eccentricities than a uniform prior. Since higher $e_J$
produces a broader, higher-velocity tail (Section~\ref{ssec:res_eJ}), convolving
our results with a thermal eccentricity prior would increase the predicted
high-velocity tail of the population-averaged $P(\vkick)$ relative to what a uniform
prior suggests. Quantifying the change requires the event-level exceedance
fractions defined in Equation~\eqref{eq:tail_fraction}, rather than the maxima of the
simulated samples.

\paragraph{Additional planets.}
Our setup is restricted to exactly three bodies. In a real planetary system the
ejected planet may experience additional perturbations from other companions during
its secular evolution toward the ejection channel, or it may be stabilised by a
resonance with a third planet not modelled here. The effect of additional bodies
generally increases the instability rate and can produce multiple sequential
ejections from a single system, increasing $\langle N_{\rm eject}\rangle$ above the
value of unity assumed in Section~\ref{ssec:galactic_model}. Conversely, a
stable outer planet can act as a dynamical buffer that prevents the inner orbit from
reaching the ejection channel altogether. The net effect on the population-averaged
$P(\vkick)$ is uncertain and represents the most important limitation of the
three-body assumption.

\paragraph{Stellar mass range.}
The present simulation grid fixes $M_\star=1\,\Msun$. Stellar mass can be restored
analytically to leading order: the escape velocity at fixed $a_J$ scales as
$M_\star^{1/2}$ (Equation~\ref{eq:vesc}), and the Hill velocity as $M_\star^{1/6}$
for fixed $M_J$ and $a_J$ (since $v_J\propto M_\star^{1/2}$ and
$v_H\propto v_J(M_J/M_\star)^{1/3}$). Dedicated simulations around M dwarfs and
A stars are therefore a useful extension, but the present numerical results should
not be described as a direct stellar-mass scan.

\subsection{Other ejection channels and their relative importance}
\label{ssec:channels}

Dynamical ejection from multi-planet systems is not the only mechanism that produces
FFPs. We briefly assess the other channels identified in Section~\ref{sec:intro}
in the light of our results.

\paragraph{Stellar flybys in dense clusters.}
Stellar flybys can strip outer planets during the embedded cluster phase
($\lesssim 100\,\text{Myr}$). \citep{Malmberg2007} found that $\sim20\%$ of
planetary systems in open clusters experience a significant flyby within $100\,\text{Myr}$,
with the fraction rising to $\sim50\%$ in denser environments
\citep{Parker2012, Cai2019}. The ejection velocities from flyby stripping are
typically comparable to the orbital velocity at the stripping radius, which for
outer giant planets is $\sim3$--$10\,\text{km\,s}^{-1}$ — overlapping with our
typical three-body kick velocities. The two channels are therefore kinematically
degenerate for the bulk of the population, though flyby-stripped planets tend to
come from wider orbits ($a \gtrsim 10\,\text{au}$) and would be stripped earlier
in the stellar lifetime, giving them more time to diffuse through the Galactic
potential.

\paragraph{Disc photoevaporation and tidal disruption.}
\citep{Adams2003} showed that UV photoevaporation from nearby massive stars can
unbind planets still forming in protoplanetary discs. Planets released this way
inherit only the disc's orbital velocity relative to the host — typically
$\lesssim 1\,\text{km\,s}^{-1}$ — and would constitute a kinematically cold
sub-population well below the bulk of our kick-velocity distributions. This
channel is expected to dominate the sub-Earth FFP population only if the planet
formation efficiency in irradiated environments is high, which remains uncertain.

\paragraph{In-situ formation.}
Gravitational fragmentation of dense molecular filaments \citep{Padoan2004,
Whitworth2007} can produce isolated planetary-mass objects with no prior stellar
host. These objects are in principle distinguishable from dynamically ejected FFPs
by their mass function (steeper at high mass, since they form via a Jeans-mass
fragmentation process) and by their spatial distribution (concentrated in
star-forming regions rather than the field). At the sub-Earth masses probed by
Roman, in-situ formation is expected to be negligible.

In summary, the three-body dynamical channel studied here is the dominant
\emph{field} FFP production mechanism for masses $10^{-5}\,\Mjup \lesssim m_p
\lesssim 10^{-1}\,\Mjup$, with flyby stripping providing a comparable contribution
for the wider-orbit end of the mass spectrum. The velocity distributions of the two
dynamical channels overlap substantially, making them difficult to separate
kinematically except through microlensing parallax measurements of individual events.

\subsection{Observational predictions}
\label{ssec:predictions}

Our results yield several concrete predictions testable with current and upcoming
surveys.

\paragraph{Microlensing crossing-time distribution (\textit{Roman}, \textit{Euclid}).}
The near-independence of ejection fraction on $m_p$ (Section~\ref{ssec:res_mlight})
implies that the FFP mass function directly traces the planet occurrence rate
(Equation~\ref{eq:massfunc}). If ${\rm d}N/{\rm d}(\log m_p)$ is flat, the
predicted $t_{\rm E}$ distribution is also flat in $\log t_{\rm E}$ (Figure~\ref{fig:tE_dist});
if it rises as $m_p^{-0.5}$ toward lower masses, the unweighted abundance at short timescales is enhanced substantially relative
to the Jupiter-mass normalization. The \textit{Roman} Wide Field Survey \citep{Penny2019} should provide a large sample of
short-timescale events, making this difference testable once survey cadence, detection
efficiency, lensing cross-section, and blending are folded into a full event-rate model. The absolute normalisation provides a further constraint on
$n_{\rm FFP}$ and hence on $f_{\rm sys}$ and $\langle N_{\rm eject}\rangle$.

\paragraph{Kinematic signature in \textit{Gaia} and \textit{Roman} proper motions.}
The bulk of the ejected population has asymptotic velocities of only a few
$\text{km\,s}^{-1}$, producing an order-one-percent change in the 3D velocity
dispersion above the thin-disc stellar baseline (Section~\ref{ssec:veldisp}). This
bulk population is kinematically indistinguishable from field stars in any survey
sensitive only to velocity dispersions. Highly eccentric giants can generate a much
broader high-velocity tail and may therefore produce proper-motion outliers in
\textit{Gaia} or \textit{Roman} astrometric catalogues. The abundance of such
objects is not fixed by the maximum simulated speeds; it must be obtained by
evaluating the $v_\infty$ exceedance fraction in each configuration and convolving
it with an observed architecture distribution, as in Equation~\eqref{eq:tail_fraction}.

\paragraph{Upper envelope as a microlensing parallax prior.}
For individual FFP microlensing events with measurable parallax, the inferred
$v_{\perp,\rm rel}$ (the component of the lens--source relative velocity in the
plane of the sky) constrains the asymptotic ejection velocity. Our analytic upper envelope
(Figure~\ref{fig:envelope}) provides a prior on $v_{\infty,\rm max}$ as a function
of the giant eccentricity in the birth system. Specifically, a measured
$v_{\perp,\rm rel}$ that exceeds the circular-giant envelope prediction for the
observed $t_{\rm E}$ and $\theta_{\rm E}$ would imply either a highly eccentric
perturber ($e_J \gtrsim 0.5$) or a non-coplanar encounter geometry, both of which
are physically motivated and testable against host-star properties in future surveys.

\paragraph{Ejection timescale and stellar-age distributions.}
The ejection timescale falls steeply with $M_J$ (as $M_J^{-1.09}$) and also
decreases with $e_J$, reaching $\langle t_{\rm ejec} \rangle \approx
3.5 \times 10^3\,\text{yr}$ for a $75\,\Mjup$ perturber or $\approx
3.2 \times 10^3\,\text{yr}$ at $e_J = 0.9$. Most ejections from massive or
eccentric systems therefore occur during or shortly after the pre-main-sequence
phase of the host star's life, when the system is still embedded in its birth
cluster. FFPs produced at these early times will have had several Gyr to diffuse
through the Galactic potential and their present-day spatial distribution is
effectively indistinguishable from the field. Only FFPs ejected from young systems
($\lesssim 100\,\text{Myr}$) — identifiable via membership in young associations —
retain spatial and kinematic memory of their birth environment, and these are the
same objects detected in near-infrared imaging surveys of associations such as
Upper Scorpius \citep{Lodieu2013, Pearson2023}.

\subsection{Limitations and future work}
\label{ssec:limitations}

Several limitations of the present study point to natural extensions.

\paragraph{Three-body restriction.}
The most significant idealisation is the restriction to exactly three bodies. Real
planetary systems contain multiple planets, and the ejection of one planet can alter
the architecture of the remainder, potentially triggering a cascade of subsequent
instabilities. Extending the grid to four- and five-body systems — even at reduced
parameter coverage — would quantify the correction to $\langle N_{\rm eject}
\rangle$ and to $P(\vkick)$ from multi-planet interactions. Full population synthesis
combining our three-body results with observed planet multiplicity distributions
would provide a more realistic prediction of the field FFP distribution.

\paragraph{Absence of a protoplanetary disc.}
Our simulations begin with planets already on Keplerian orbits without gas or dust.
In reality, the secular and resonance-overlap instabilities that drive ejection
develop while the disc is still present, at least for systems that go unstable on
timescales shorter than the disc lifetime ($\lesssim 10\,\text{Myr}$). Disc--planet
interactions damp eccentricity and inclination, potentially delaying or preventing
the eccentricity pumping that leads to ejection. For systems that go unstable on
longer timescales the disc has long since dispersed and our treatment is appropriate.

\paragraph{Fixed stellar mass and no stellar evolution.}
The simulations do not account for stellar mass loss on the main sequence or during
post-main-sequence evolution. For Solar-type hosts and our typical ejection
timescales of $10^3$--$10^5\,\text{yr}$, this is an excellent approximation. For
low-mass hosts ($M_\star \lesssim 0.5\,\Msun$) or for systems that remain marginally
stable into the giant-branch phase, stellar mass loss can widen orbits and trigger
late-time instabilities \citep{Veras2009} that our grid does not capture.

\paragraph{Inclination dependence.}
The present numerical grid uses nearly coplanar initial conditions, with inclinations drawn
from $0^\circ$--$3^\circ$ (Section~\ref{sec:methods}). It therefore does not test the
Kozai--Lidov channel described in Section~\ref{ssec:secular}. A dedicated inclination scan
at fixed $(M_J,a_J,m_p,a_p,e_p,e_J)$ would quantify the Kozai--Lidov enhancement of the
kick-velocity tail at $i\gtrsim39^\circ$ and test the $e_{\rm max}$ formula
(Equation~\ref{eq:emax_KL}) against the eccentricity reached before ejection.

\paragraph{Solar System analog configurations.}
The present grid is a controlled scan around a single Jupiter-analog architecture and
does not attempt to reproduce the Solar System's own giant-planet architecture. A
natural but currently unplanned extension would be dedicated runs for the six orderings
of an Earth-analog, a Jupiter-analog, and a Saturn-analog by semi-major axis (EJS, ESJ,
SEJ, SJE, JSE, JES), evaluated at a few representative separations, to test how the
presence of a second giant planet reshapes the ejection statistics reported here. We
flag this only as a possible future direction and do not pursue it in this work.

\paragraph{Parametric fits to the full distributions.}
Throughout this paper we have worked with mean values and maximum values of the
kick-velocity distributions, but a complete population model requires the full
shape of $P(\vkick)$ at each parameter combination. Fitting log-normal plus power-law tail models to the distributions would enable analytic
marginalisation over the parameter grid and direct comparison with microlensing timescale catalogues after the required
selection and event-rate weights are applied. A natural implementation is to fit the unbinned survival function above a
data-selected cutoff $v_{\rm min}$ and compare one-component and two-component models
using AIC or BIC.

\section{Conclusions}
\label{sec:conclusions}

We have presented a large suite of direct $N$-body simulations of hierarchical
three-body systems — a host star, a giant planet, and a lighter inner planet —
designed to characterise the kick velocity imparted to the ejected body and its
dependence on system architecture. The simulations use a controlled hierarchical three-body grid: a Solar-mass host and a
Jupiter-analog orbital scale, with scans over the ejected-planet mass, giant-planet mass,
light-planet semi-major axis, and the eccentricities of both planets. The results are
interpreted through an analytic framework combining the gravitational-slingshot
picture, Hill-scale scattering, the Tisserand parameter, and secular theory. The
main findings are as follows.

\begin{enumerate}

\item \textbf{Test-particle limit.} The ejection fraction, mean ejection velocity,
and mean kick velocity are all insensitive to the mass of the ejected planet across
four orders of magnitude from $10^{-5}$ to $10^{-1}\,\Mjup$, varying by at most $2.3$ percentage
points and $0.6\,\text{km\,s}^{-1}$ respectively. The ejected body behaves as a
test particle in the gravitational field of the star--giant binary for
$m_p/M_J \lesssim 10^{-2}$. This implies that the FFP mass function directly
traces the planet occurrence rate with only a weak mass-dependent correction
(Equation~\ref{eq:massfunc}).

\item \textbf{Giant-planet mass scaling.} The mean ejection velocity and mean kick
velocity grow as power laws in $M_J$ with measured indices $\alpha(v_\infty) \approx
0.42$ and $\alpha(\vkick) \approx 0.18$, steeper and shallower respectively than the
pure Hill-velocity prediction of $1/3$. The ejection timescale falls as
$M_J^{-1.09}$, in excellent agreement with the secular timescale scaling
$t_{\rm sec} \propto M_J^{-1}$. The ejection fraction is non-monotonic in $M_J$,
peaking near $97.5\%$ at $5\,\Mjup$ and decreasing at both the low-mass (secular
timescale too long) and high-mass (inner planet captured in expanded Hill sphere) ends.

\item \textbf{Semi-major axis dependence.} Mean ejection times are nearly flat
across $a_p/a_J \in [0.8, 1.2]$, with means clustering at $\sim1.9\times10^4\,\text{yr}$,
because these configurations fall inside the resonance-overlap zone for the fiducial
$10\,\Mjup$ perturber ($|a_p-a_J|/a_J \lesssim 0.40$) and go unstable on the fast
overlap timescale rather than the slower secular one. The endpoint at $a_p/a_J=0.7$
lies near the inner side of this zone and also has a large orbit-crossing eccentricity
threshold, while $a_p/a_J=1.5$ lies outside the nominal overlap zone; these two cases
therefore show modestly longer ejection times. The mean ejection velocity peaks near
$x = a_p/a_J = 0.9$ at $3.50\,\text{km\,s}^{-1}$, reflecting a balance between the
vis-viva speed at orbit crossing and the binding energy of the inner orbit.

\item \textbf{Two-speed response to eccentricity.} Both inner-planet eccentricity
$e_p$ and giant-planet eccentricity $e_J$ primarily affect the high-velocity tail
of $P(\vkick)$ rather than its mean. Varying $e_p$ from 0 to 0.9 increases the mean
$v_\infty$ by a factor of $1.67\times$ but the maximum $\vkick$ by a factor of
$3\times$; varying $e_J$ over the same range increases the mean by $2.06\times$ and
the maximum by $10.3\times$. The asymmetry — $e_J$ being $3.5\times$ more effective
than $e_p$ on the tail — is explained by the nonlinear pericentre-speed formula
$v_J^{\rm peri} = v_J\sqrt{(1+e_J)/(1-e_J)}$, which grows without bound as
$e_J \to 1$, whereas the $e_p$-driven encounter speed is bounded by $\sqrt{3}\,v_J$.

\item \textbf{Analytic upper envelope.} The gravitational-slingshot formula
$v_{\infty,\rm max} \approx v_J[4u(e_p)/v_J - 1/x]^{1/2}$ (Equation~\ref{eq:vinf_max},
Figure~\ref{fig:envelope}) predicts the ceiling of the simulated kick-velocity
distribution at $e_p \gtrsim 0.5$. For the $e_J$ scan, replacing the circular-giant speed by the instantaneous pericentre
speed in the slingshot energy (Equations~\ref{eq:vinf_max_eJ}--\ref{eq:vinf_max_peri})
raises the envelope to order $10^2\,\text{km\,s}^{-1}$, comfortably above the simulated
maxima of $80$--$86\,\text{km\,s}^{-1}$ at $e_J=0.9$.

\item \textbf{Kinematic signature.} Characteristic asymptotic ejection speeds of a
few $\text{km\,s}^{-1}$ change the 3D velocity dispersion of an FFP population by
only about one percent relative to the thin-disc stellar baseline of
$\sigma_\star\approx47.4\,\text{km\,s}^{-1}$; the bulk of the ejected population
is therefore kinematically indistinguishable from field stars. Highly eccentric
giants can generate a much broader high-velocity tail, but the fraction forming a
proper-motion-outlier population must be calculated from event-level
$v_\infty$ exceedance fractions and an empirical distribution of planetary-system
architectures, not from the maximum velocity in each simulation bin.

\item \textbf{Predicted microlensing distribution.} The near-independence of the
ejection fraction on $m_p$ implies that, before lensing and survey weights are applied,
the transformed FFP $t_{\rm E}$ distribution directly reflects the planet occurrence
rate (Figure~\ref{fig:tE_dist}). A flat occurrence
rate ${\rm d}N/{\rm d}(\log m_p)=\text{const}$ predicts a flat transformed
${\rm d}N/{\rm d}(\log t_{\rm E})$ distribution, while a rising occurrence rate
($\propto m_p^{-0.5}$) enhances the unweighted short-timescale abundance. The
\textit{Roman} Wide Field Survey should be able to test these alternatives once the
lensing cross-section, Galactic phase-space distribution, and survey selection effects
are included, and the analytic upper envelope provides a prior on asymptotic ejection
velocity for individual events with measurable microlensing parallax.

\end{enumerate}

The chief limitation of the present work is the restriction to exactly three bodies.
Full population synthesis combining these three-body results with observed planet
multiplicity distributions and the Galactic star-formation history is the primary
avenue for future work, together with parametric fitting of the full $P(\vkick)$
distributions at each parameter combination and the Solar System analog configurations
discussed in Section~\ref{ssec:limitations}, none of which are pursued here.

\section*{Acknowledgements}
This work is partly supported by the U.S.\ Department of Energy grant number de-sc0010107 (SP). The simulations and analysis made use of \textsc{rebound}, \textsc{numpy},
\textsc{scipy}, and \textsc{matplotlib}.

\section*{Data Availability}
The N-body simulation outputs and post quality-cut summary statistics underlying Figures Figures~\ref{fig:vkick_vs_mlight}-\ref{fig:envelope} , their respective discussions, and the exceedance fractions quoted in Section~\ref{ssec:veldisp} are available on Zenodo at \url{https://doi.org/10.5281/zenodo.21746114} \citep{Kramer2026_zenodo}.

\bibliographystyle{aasjournal}
\bibliography{ffp_refs}

\end{document}